\documentclass[journal]{IEEEtran}

\usepackage[
    font=footnotesize,          % match IEEE style
    justification=centering,    % center multi-line captions
    singlelinecheck=false       % also center single-line captions
]{caption}

\usepackage{float}
\usepackage{amssymb}
\usepackage{cite} 
\usepackage[colorlinks=true,linkcolor=black,citecolor=green,urlcolor=blue]{hyperref}
\usepackage{xcolor}

\ifCLASSINFOpdf
\usepackage[pdftex]{graphicx}
\graphicspath{{../pdf/}{../jpeg/}}
\DeclareGraphicsExtensions{.pdf,.jpeg,.png}
\else
\fi
\usepackage{adjustbox}  % loads graphicx internally

\usepackage{amsmath}
\usepackage{graphicx}
\usepackage{balance}
\usepackage{listings}

\begin{document}
\title{HFEMCNet: A Compact Hybrid Frequency Enriched Multi Channel Network for Automatic Modulation Classification}

\author{Qamar~Ijaz, %,~\IEEEmembership{Member,~IEEE,}
        Nayyer~Aafaq % ,~\IEEEmembership{Fellow,~OSA,}
       % and Other~,~\IEEEmembership{Life~Fellow,~IEEE}% <-this % stops a space
% \thanks{Q. Ijaz and N. Aafaq is with the School of Avionics and Electrical Engineering, College of Aeronautical Engineering, NUST, Risalpur, Pakistan, e-mail: (see qijaz.ms17avecae@student.nust.edu.pk, naafaq@cae.nust.edu.pk).}% <-this % stops a space
\thanks{}% <-this % stops a space
\thanks{}}

\maketitle

\begin{abstract}
Automatic modulation classification (AMC) of received radio signals is prudent for further signal processing tasks such as communication monitoring, cognitive radio operation, and interference mitigation in the electromagnetic spectrum. Traditional methods often rely on handcrafted features and struggle under complex channel conditions, whereas deep learning (DL) architectures can learn discriminative representations directly from raw signals. In this work, we propose HFEMCNet, a novel compact hybrid frequency-enriched multi-channel network that jointly exploits spatiotemporal dependencies and frequency-domain information derived via Fast Fourier Transform (FFT). HFEMCNet integrates raw IQ samples with hierarchical spectral features to produce a rich signal representation, which is processed by convolutional layers for spatial feature extraction and Long Short-Term Memory (LSTM) units for temporal modeling. Experimental evaluation on benchmark datasets RML2016.10a, RML2016.10b, and over-the-air RML2018.01a demonstrates that HFEMCNet significantly outperform contemporary state-of-the-art DL models in classification accuracy while achieving a reduced parameter count, smaller memory footprint, and lower tail-latency, making it suitable for real-time deployment on resource-constrained platforms. These results highlight the effectiveness of combining hybrid architectures with frequency-domain enrichment for robust and efficient AMC.
\end{abstract}


% \fi
\IEEEpeerreviewmaketitle

\section{Introduction}

\IEEEPARstart{E}{M} spectrum forms the backbone of modern wireless communications, underpinning a broad spectrum of wireless services, including mobile networks, radar systems, satellite links, and emerging autonomous systems. 
Efficient and secure utilization of this spectrum requires unprecedented awareness of the transmitted signals, with accurate identification of modulation schemes serving as a cornerstone for reliable communication~\cite{source1}. 
Deep Learning-based Automatic Modulation Classification (DL-AMC) has emerged as paradigm, enabling modulation classification without any prior information of the radio signals. In practical scenarios, transmitted radio signals are inevitably distorted due to additive noise, channel impairments, multi-path fading, and carrier frequency offsets~\cite{source2} resulting in received signals that deviate significantly from their original form. 
Accurate demodulation at the receiver therefore necessitates precise knowledge of the modulation scheme.

Traditional AMC approaches can be broadly categorized into likelihood based and feature based methods~\cite{source3}. 
Likelihood-based AMC (LB-AMC), employs multiple hypothesis testing to compute the likelihood of each candidate modulation, providing theoretically optimal performance under ideal conditions.
However, LB-AMC is often computationally expensive requiring precise channel state information limiting its practicality in real-world deployments~\cite{source4}. 
Feature-based AMC (FB-AMC), on the other hand, relies on manually engineered features, such as Higher Order Moments (HOMs) and cumulants, to perform classification. 
While FB-AMC offers reduced complexity and improved classification performance, its effectiveness is highly dependent on expert domain knowledge for feature extraction and the discriminative quality of the selected features~\cite{source5}.

\begin{figure*}[t]
    \centering
    \includegraphics[width=0.9\linewidth]{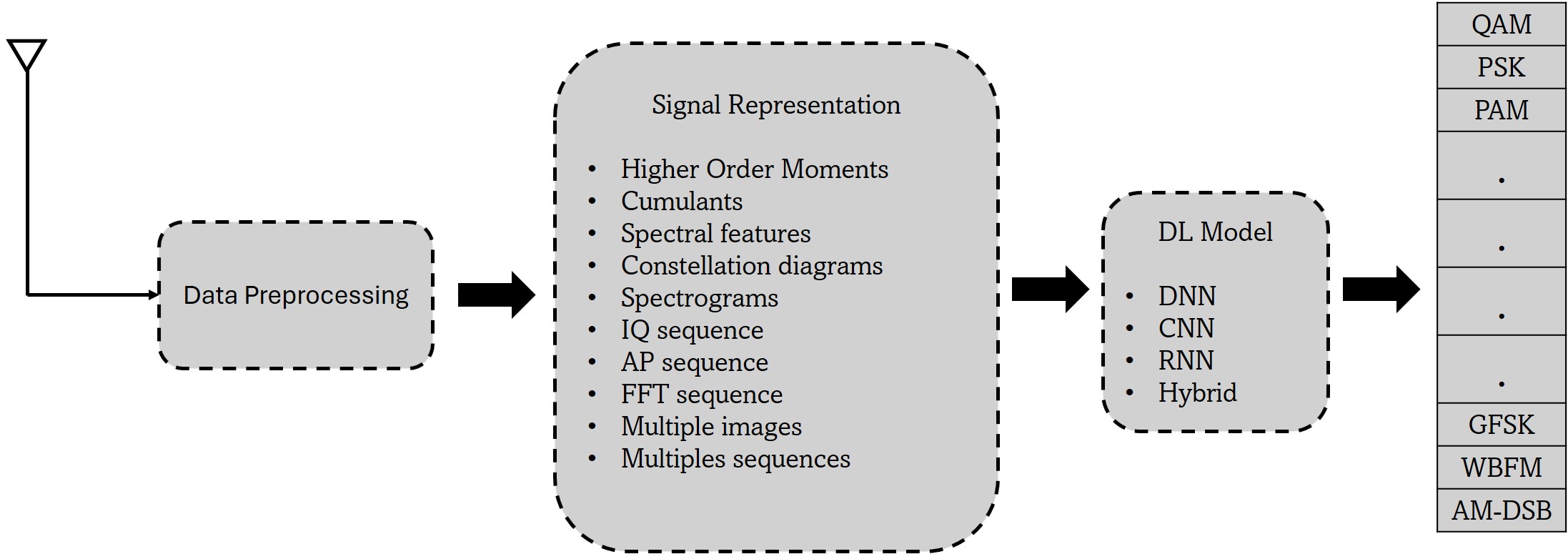}
    \caption{Overview of the DL-AMC workflow, illustrating the three main stages: signal pre-processing, signal representation, and deep neural network–based classification. The diagram also highlights signal representation techniques (image-based, sequence-based, feature-based, hybrid etc.) and the corresponding deep learning model categories leading to modulation scheme identification}
    \label{fig:categories}
\end{figure*}

Deep learning based AMC approaches have demonstrated superior performance compared to traditional methods, driven by the availability of benchmark datasets, advances in deep neural network architectures, and high-performance computing platforms. 
The DL-AMC workflows can be broadly divided into three stages: signal pre-processing, signal representation and deep neural network for classification as illustrated in Fig.~\ref{fig:categories}. Based on the type of signal representation, DL-AMC methods can be further categorized into four classes namely image-based, sequence-based, feature-based, and hybrid representations~\cite{source6}.  
Each representation paradigm offers distinct advantages and limitations, which can be exploited to achieve optimal classification performance under diverse channel conditions.

%----------------------------------------------------------------------------
%----------------------------------------------------------------------------

In this paper, we propose a compact Hybrid Frequency-Enriched Multi-Channel Network (HFEMCNet) for Automatic Modulation Classification. 
The proposed architecture addresses the limitations of contemporary modulation classification techniques by simultaneously exploiting spatial, temporal, and frequency-domain features within a unified DL framework. 
Specifically, HFEMCNet adopts a multi-channel structure, where separate convolutional neural networks (CNNs) extract spatial features from heterogeneous signal representations, learning hierarchical spatial descriptors from the raw in-phase (I) and quadrature (Q) components as well as from their corresponding Fast Fourier Transform (FFT) magnitude spectra. 
Subsequently, temporal dependencies are modeled through Long Short-Term Memory (LSTM) layers, whose gated recurrent dynamics capture long-range inter-symbol correlations and non-stationary temporal statistics, thereby enhancing discrimination of complex modulation patterns over time.
%%%%%%%%%%%%%%%%%%%%%%%%%%%%%%%%%%%%%%%%%%%%%%%%%%%%%%%%%%%%%%%%%%%%%%%%%%%%%%%%%%%%%%%%%%%

Based upon the proposed framework, the main contributions of the proposed method are summarized as follows:
\begin{itemize}

    \item Proposed HFEMCNet architecture employs distinct yet interconnected pathways that independently process raw I/Q components and their corresponding FFT magnitudes, followed by a dedicated feature fusion mechanism. This approach not only ensures comprehensive signal representation across both time and frequency domains but also significantly reduces model complexity compared to contemporary state-of-the-art DL-AMC methods.
    
    \item We leverage Gaussian Dropout layers to reinforce the generalization capability of the model, making it robust across diverse signal-to-noise ratio (SNR) conditions and channel impairments. We incorporate Gaussian Dropout layers across the multi-channel pathways, mitigating overfitting and improving generalization across diverse SNR conditions. Unlike conventional AMC models that apply dropout uniformly, our approach leverages pathway-specific stochastic regularization, ensuring both temporal and spectral feature streams are resilient to noise and channel distortions, thereby representing a novel regularization strategy in AMC architectures.
    
    \item We validate our approach through extensive experiments on three benchmark datasets RML2016.10a, RML2016.10b, and RML2018.01a demonstrating superior performance of HFEMCNet. The proposed method achieves higher classification performance, outperforming contemporary AMC models. The results confirm the model’s robustness across diverse modulation schemes and SNR conditions.
    
    \item In addition to higher classification accuracy, the proposed model maintains a compact parametric footprint and computationally efficient architecture. The architectural efficiency ensures low-latency inference rendering the framework well-suited for real-time and resource-constrained scenarios. 
    
\end{itemize}

\section{Related Work}

In this section, we discuss various deep learning (DL) architectures employed for modulation classification. For clarity, the DL models are grouped according to their baseline architectures. DL-based automatic modulation classification (AMC) generally consists of three phases, as illustrated in Fig.~\ref{fig:categories}. First, data pre-processing is performed to prepare the signals for representation and subsequent training. Next, signal representation transforms the data to enrich and align with the input requirements of the DL model. Finally, the DL model is trained to perform modulation classification.

\subsection{CNN Based Models}\label{cnnmodels}

Convolutional neural network (CNN)-based deep learning frameworks have demonstrated superior classification performance due to their ability to effectively capture spatial characteristics in the data. Authors in~\cite{source2} and~\cite{source11} proposed a simple four-layer CNN architecture using raw IQ samples as input. These architectures outperformed traditional AMC methods on real-world complex datasets; however, classification errors between higher-order modulation schemes, such as 16-QAM and 64-QAM, remained significant even at high signal-to-noise ratio (SNR) ranges. To address this, authors in~\cite{source12} introduced a two-stage CNN framework: the first CNN is trained on IQ data for initial classification, while a second CNN trained on constellation diagrams refines classification of higher-order modulations. This approach achieved improved accuracy for complex modulation schemes.

In~\cite{source13}, the basic CNN architecture of~\cite{source2} was extended with deeper layers, and adaptive noise was introduced during training to enhance noise robustness. While this architecture is more complex, it is resilient to channel impairments and suitable for real-world scenarios. In~\cite{source14}, a novel architecture based on residual connections was proposed to mitigate the vanishing gradient problem, enabling the network to learn complex modulation patterns. The same study also introduced a dense network framework, connecting each layer to every other layer, which proved highly effective for extracting intricate features from complex datasets. Authors in~\cite{source15} proposed IC-AMCNet, a CNN-based model for automatic modulation classification in beyond-5G communications, incorporating dropout and Gaussian noise layers to reduce overfitting and improve latency. Additionally, a novel DL-AMC model in~\cite{source16} addressed channel impairments in wireless communications by employing asymmetric convolutional kernels and skip connections, effectively mitigating gradient vanishing and achieving higher classification accuracy than contemporary DL-AMC models. Further, in~\cite{source17}, a 1D CNN parallel fusion framework was introduced, which is computationally efficient and achieves improved classification performance. The proposed DL-AMC model demonstrates consistently improved classification performance compared to standard CNN architectures, while simultaneously providing enhanced robustness and better discrimination of higher order modulation schemes under varying SNR conditions.

%----------------------------------------------------------------------------
\begin{figure}[t]
    \centering
    \includegraphics[width=0.9\linewidth]{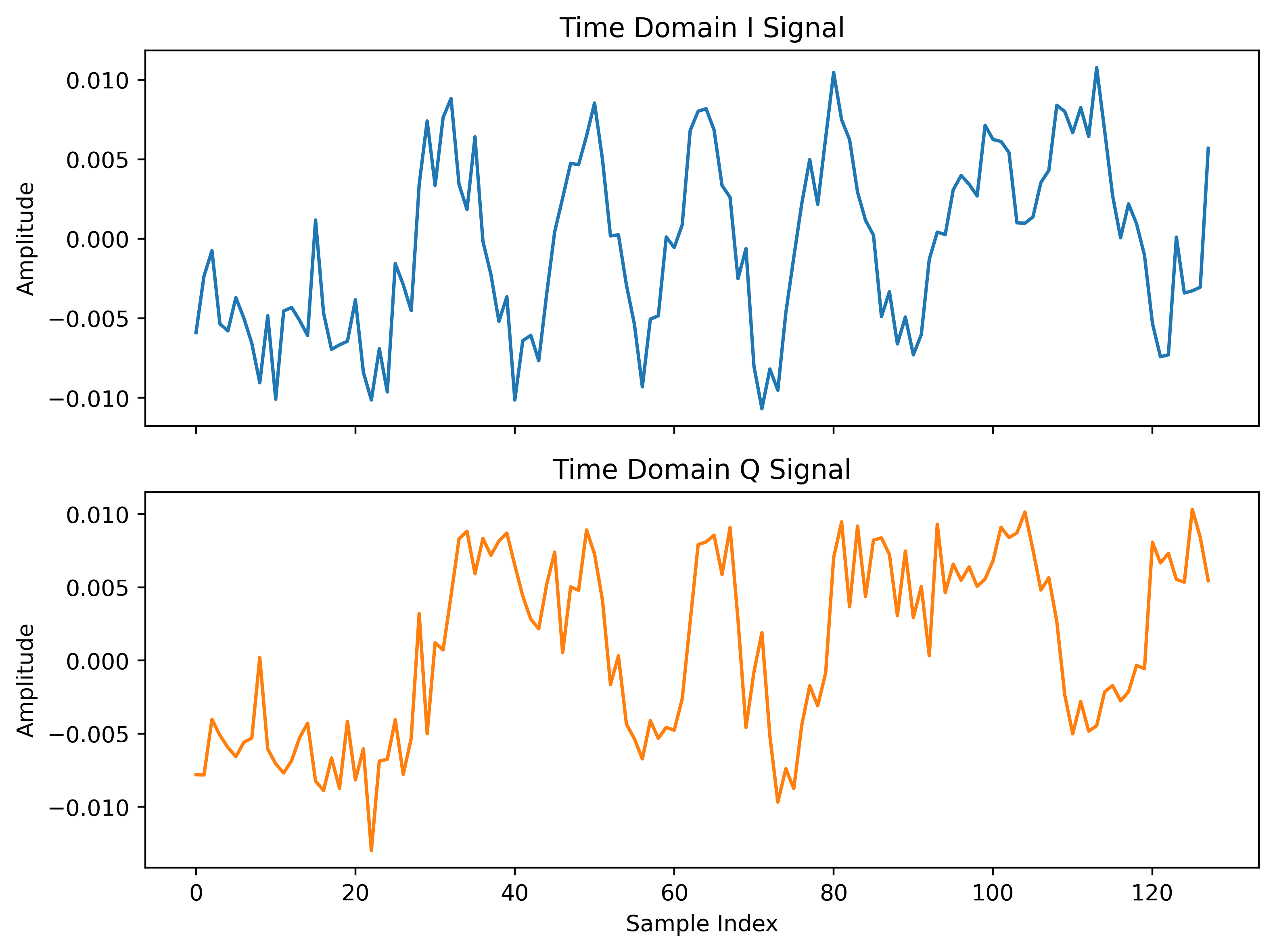}
    \caption{Depiction of time-domain in-phase (I) and quadrature (Q) Signals}
    \label{fig:rawsample}
\end{figure}
%----------------------------------------------------------------------------

\subsection{Temporal Modeling with RNNs}\label{rnn}

As CNN based DL-AMC is able to characterize the spatial features, however,  temporal dependencies can be  learned using Recursive Neural Networks (RNNs) to classify the modulation type of radio signals. Authors in \cite{source18} have used Gated Recurrent Units (GRU), a variant of RNN, to exploit the temporal sequence characteristics of radio signals. In the proposed architecture two layers of GRUs are used with IQ samples as input. The two layers GRU architecture was able to efficiently extract the temporal dependencies. In \cite{source19}, LSTM based DL model using amplitude and phase information in time domain is proposed. Time domain amplitude and phase information is extracted from IQ signals. Proposed architecture is adaptive with respect to handle different symbol rate and SNR condition.  Proposed LSTM based model has performed exceptionally well across different datasets. Overall classification performance of the model is significantly higher than the CNN based DL-AMC models. RNNs based DL automatic modulation classification have been extensively employed in the hybrid DL architectures. In \cite{source20}, RNN based LSTM auto encoder is introduced to enhance modulation classification in complex and noisy environment.  LSTM is used to capture temporal dependencies and denoising auto encoder helps learn low dimensional features.

\subsection{Hybrid Models}\label{hybridmodels}

 CNN and RNN based DL model for automatic modulation classification have demonstrated promising performance, particularly in complex environments with noise and channel impairments. To further enhance the modulation classification accuracy, various hybrid architectures combining CNN and RNN have been introduced. In~\cite{source21}, the authors proposed a hybrid AMC architecture to balance high recognition accuracy with low model complexity. The architecture includes a parameter estimator and transformer module to estimate phase offsets for phase compensation. A hybrid model consisting of CNN and GRU layers is employed for features extraction. CNN is used to capture spatial features while GRU is used for extraction of temporal dependencies.
 
%----------------------------------------------------------------------------
 \begin{figure}[t]
    \centering
    \includegraphics[width=0.9\linewidth]{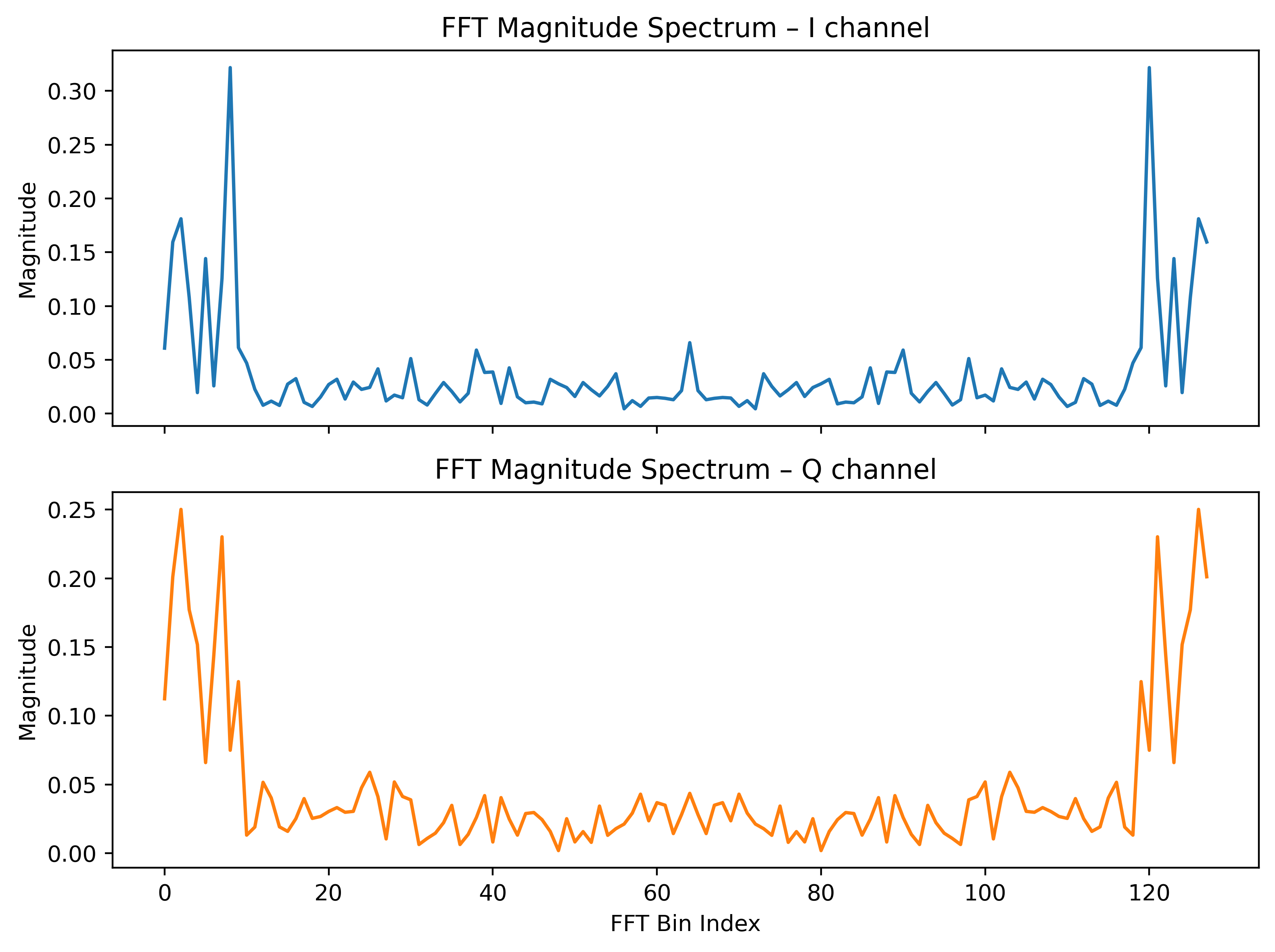}
    \caption{FFT Magnitude Spectrum of Corresponding I/Q Signals}
    \label{fig:fftmagnitude}
\end{figure}
%---------------------------------------------------------------------------- 
  
The hybrid architecture CGDNet~\cite{source22} integrates CNN layers for spatial features extraction,  GRU layers for temporal correlation, and a DNN for the final classification, with Gaussian dropout and skip connections addressing the vanishing gradient and overfitting issues. 
Similarly, in~\cite{source23}, a hybrid model with two CNN layers and two LSTM layers, namely convolutional LSTM deep neural network (CLDNN), captures both spatial and temporal dependencies. 
In~\cite{source14} and~\cite{source23}, it has been demonstrated that combining CNNs with LSTMs can significantly enhance the performance of modulation recognition in challenging EM environments. 
The idea of combining CNN and RNN based architecture has been further investigated with IQ data being fed in form of different channel to exploit the spatiotemporal characteristics of the input data. IQ data has been split into three channels namely I channel, Q channel and combined IQ channel into three respective CNN layers. After features fusion from these layers, two LSTM layers have been used to find the temporal dependencies. Therefore, a multi-channel hybrid network has been formulated known as MCLDNN~\cite{source24}. The MCLDNN model has shown improved classification performance when compared with contemporary DL-AMC models~\cite{source32}~\cite{source33}.

In \cite{source25}, a hybrid model combining ResNeXt and GRU was proposed, where ResNeXt captures spatial features and GRU models temporal dependencies. Then extracted features are then fused using Discrimination Correlation Analysis (DCA). In~\cite{source26}, a Hybrid Knowledge and Data Driven (HKDD) DL-AMC model has combined the spatial and handcrafted features using attention mechanism to enhance modulation classification performance.

Authors in~\cite{source27}, introduced the prior regularization in the form of inter-class antagonistic factor, as well as global and dimensional divergence, preceding the hybrid CNN-LSTM model. Incorporation of this prior regularization significantly improved the classification performance compared to standalone CNN-LSTM models, particularly in the low SNR region which is considered to be a challenging area. Overall, the accuracy in the low SNR region has been progressively enhanced by exploiting spatiotemporal features through hybrid CNN-RNN models and efficiently fusing these features using strategies such as spatiotemporal decoders and attention mechanisms~\cite{source28}~\cite{source29}.

\section{Radio Signal Model/ Representation}
\subsection{Radio Signal Model}
At receiver end, the transmitted signal undergoes multiple propagation-induced distortions, primarily due to multipath fading and additive noise. These impairments introduces both structural and stochastic alterations to the received signal. For a single input single output (SISO) communication system, the received signal \(y(t)\) can be modeled as:
%-------------------------------------------------
\begin{equation}
y(t) = x(t)*h(t) + i(t)
\label{eq:sigmod}
\end{equation}
%-------------------------------------------------
where \(x(t)\) denotes the transmitted signal, \(h(t)\) represents the channel impulse response, \(i(t)\) denotes additive noise, and \(*\) indicates the convolution operation. Convolution of transmitted signal \(x(t)\) with multipath effect denoted by \(h(t)\) represents that the transmitted signal reaches at receiver with different time delays. Moreover, \(i(t)\) represents the addition of noise. Multipath effect \(h(t)\) can be represented by:
%-------------------------------------------------
\begin{equation}
 h(t) = \sum_{n=1}^{N} \ a_n e^{j\phi_n} \delta(t - \tau_n)
\end{equation}
%-------------------------------------------------
where $a_n$ denotes the complex gain of the $n^{\text{th}}$ propagation path, capturing amplitude attenuation due to large-scale fading and small-scale channel impairments. The term $e^{j\phi_n}$ represents the phase shift introduced by the $n^{\text{th}}$ path as a result of propagation-induced delay and carrier frequency offset. The delay associated with each multipath component is denoted by $\tau_n$, and $N$ indicates the total number of resolvable propagation paths. Accordingly, the received signal incorporating multipath effects can be modeled as:

%-------------------------------------------------
\begin{equation}
y(t) = \sum_{n=1}^{N} \;a_n e^{j\phi_n} \;x(t - \tau_n) + i(t)
\end{equation}
%-------------------------------------------------

Mixing the received passband signal $y(t)$ with coherent local oscillator signals $\cos(2\pi f_c t)$ and $\sin(2\pi f_c t)$, followed by ideal low-pass filtering (LPF), yields the complex baseband representation. The resulting in-phase and quadrature components are denoted by $y_I(t)$ and $y_Q(t)$, respectively. The equivalent complex baseband signal is therefore expressed as:

%-------------------------------------------------
\begin{equation}
y_{\mathrm{bb}}(t) = y_I(t) + y_Q(t)
\end{equation}
%-------------------------------------------------
The complex baseband signal $y_{\mathrm{bb}}(t)$ is subsequently digitized using an analog-to-digital converter (ADC), resulting in the discrete-time sequence $y[n]$. Uniform sampling is performed at instants $t_n = nT_s$, where $T_s$ denotes the sampling interval. The corresponding sampling frequency is defined as $f_s = 1/T_s$.
The resulting discrete-time complex baseband sample at index $n$ is expressed as
%-------------------------------------------------
\begin{equation}
y[n] = I[n] + j Q[n],
\end{equation}
%-------------------------------------------------
where $I[n]$ and $Q[n]$ denote the in-phase and quadrature components, respectively.
where $n = 0,1,2,\dots,N-1$, and $I[n]$ and $Q[n]$ denote the in-phase and quadrature components of the sampled received signal at instant $n$. The imaginary unit is defined as $j = \sqrt{-1}$. These I/Q components are employed as features for training and testing the deep learning model. For each observation, the I/Q samples can be arranged in a matrix of dimension $2 \times N$, where $N$ corresponds to the total number of discrete-time samples collected for a single example. Each sample is represented as

\begin{equation}
\mathbf{y} =
\begin{bmatrix}
\mathbf{I}[n] \\[2pt]
\mathbf{Q}[n]
\end{bmatrix}
=
\begin{bmatrix}
I[0] & I[1] & \dots & I[N-1] \\[2pt]
Q[0] & Q[1] & \dots & Q[N-1]
\end{bmatrix}.
\end{equation}
% %%%%%%%%%%%%%%%%%%%%%%%%%%%%%%%%%%%%%%%%%%%%%%%%%%%%%%%%

Benchmark datasets employed for evaluating our model store raw I/Q samples in matrices of dimension $2 \times N$, where one row corresponds to the in-phase component and the other to the quadrature component. Each observation thus contains $N$ discrete-time values per I and Q vector. For instance, the RML2016 dataset provides $N = 128$ samples per I/Q vector, whereas the RML2018 dataset contains $N = 1024$ samples per vector.

\subsection{Signal Representation}

We first compute the Fast Fourier Transform (FFT) of length $N$ for the raw in-phase ($I[n]$) and quadrature ($Q[n]$) vectors from the dataset. The FFT is an efficient implementation of the discrete Fourier transform (DFT), which converts a discrete-time, finite-duration sequence into the frequency domain. Each frequency bin represents the correlation between the input sequence and a complex sinusoid at a specific frequency. The DFT is defined as
%-------------------------------------------------
\begin{equation}
B_{k} \;=\; \sum_{n=0}^{N-1} e^{-\,i \,\frac{2\pi}{N}\,k\,n}\;\mathbf{y}_{n}
\end{equation}
%-------------------------------------------------

%----------------------------------------------------------------------------
\begin{figure}[t]
    \centering
    \includegraphics[width=0.85\linewidth]{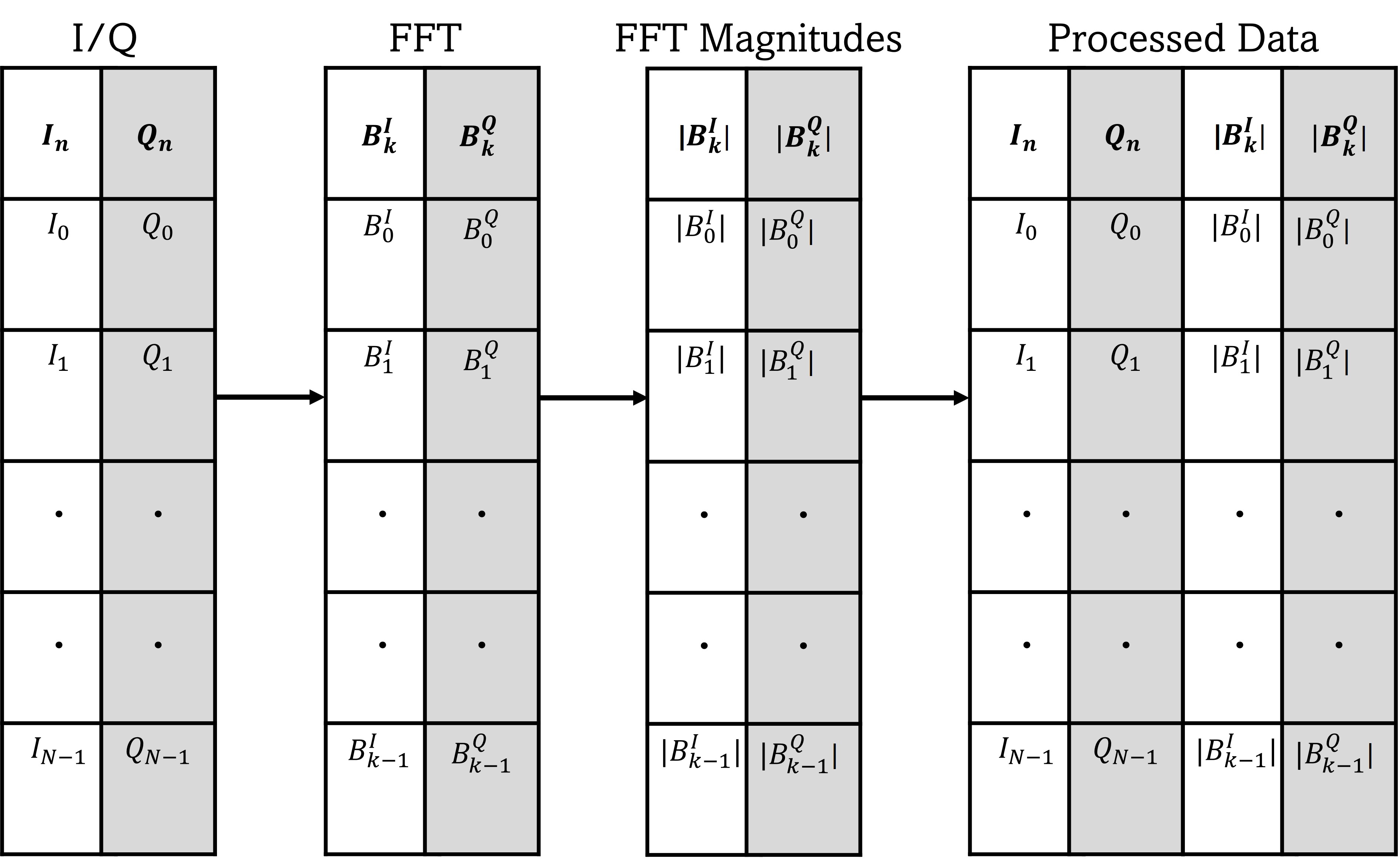}
    \caption{Data Preprocessing Breakdown}
    \label{fig:dataprocessing}
\end{figure}
%----------------------------------------------------------------------------
where $B_k$ represents the $k$th frequency-bin sequence, and $\mathbf{y}_n$ is the input time-domain sequence, corresponding to either the $\mathbf{I}_n$ or $\mathbf{Q}_n$ vector in our signal representation. The term $e^{-i \frac{2\pi}{N} k n}$ is the complex exponential correlation function, while $n$ and $k$ denote the input and output indices, respectively. The DFT generates $N$ frequency bins, equal to the length of the input sequence. By summing the time-domain sequence multiplied by the correlation function, we compute the frequency component in each bin. The complex exponential $e^{-i \frac{2\pi}{N}}$ is commonly denoted as $X_N$, referred to as the twiddle factor. Thus, the DFT can be expressed as:

\begin{equation}
\mathbf{B}_{k} \;=\; \sum_{n=0}^{N-1} X_N^{\,k\,n}\;\mathbf{y}_{n}
\end{equation}

For each value of \(k\) and \(n\), the matrix form of DFT is represented by:

\begin{equation}
\resizebox{0.91\columnwidth}{!}{% <-- 95% of column width, leaves 5% for the number
  $\displaystyle
  \begin{bmatrix}
    B_0 \\[4pt]
    B_1 \\[4pt]
    B_2 \\[4pt]
    \vdots \\[4pt]
    B_{N-1}\\[4pt]
  \end{bmatrix}
  =
  \begin{bmatrix}
    X^{0\cdot0} & X^{0\cdot1} & \cdot\cdot & X^{0\cdot(N-1)} \\[4pt]
    X^{1\cdot0} & X^{1\cdot1} & \cdot\cdot & X^{1\cdot(N-1)} \\[4pt]
    X^{2\cdot0} & X^{2\cdot1} & \cdot\cdot & X^{2\cdot(N-1)} \\[4pt]
    \vdots  & \vdots  & \vdots & \vdots      \\[4pt]
    X^{(N-1)\cdot0} & X^{(N-1)\cdot1} & \cdot\cdot & X^{(N-1)\cdot(N-1)}\\[4pt]
  \end{bmatrix}
  \begin{bmatrix}
    y_0 \\[4pt]
    y_1 \\[4pt]
    y_2 \\[4pt]
    \vdots \\[4pt]
    y_{N-1}\\[4pt]
  \end{bmatrix}$
}
\label{eq:dft-matrix}
\end{equation}

%----------------------------------------------------------------------------
\begin{figure*}[t]
    \centering
    \includegraphics[width=1\linewidth]{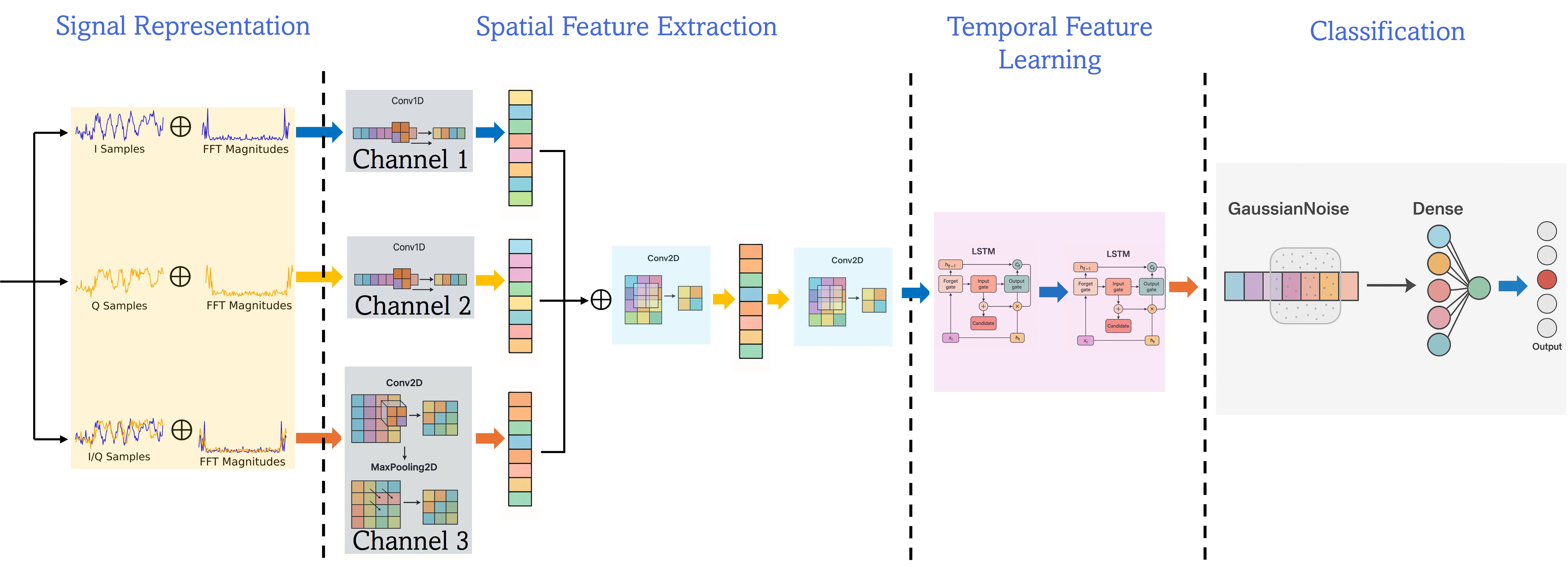}
    \caption{Proposed lightweight Hybrid Frequency-Enriched Multi-Channel Network (HFEMCNet). HFEMCNet  architecture comprises three parallel channels capturing complementary signal characteristics. The \textit{first} and \textit{second channels} process the in-phase ($\mathbf{I}$) and quadrature ($\mathbf{Q}$) sequences, respectively, along with their FFT magnitude vectors, while the \textit{third channel} jointly processes combined I/Q sequences and their FFT magnitudes to learn cross-component dependencies. Each channel comprises \textit{three convolutional layers} for feature extraction, followed by an \textit{LSTM layer} to model long-term \textit{temporal dynamics}}
    \label{fig:framework}
\end{figure*}
%----------------------------------------------------------------------------

As the sequence length grows to thousands or millions, the computational cost of the DFT increases significantly. To address this, we utilize the Fast Fourier Transform (FFT) to efficiently compute the DFT by exploiting repeated patterns within the transform. The FFT reduces the computational complexity from $O(N^2)$ for the DFT to $O(N \log N)$. Accordingly, computing the FFT of the in-phase ($\mathbf{I}$) and quadrature ($\mathbf{Q}$) vectors produces $k$ frequency bins, each represented as a complex number:

\begin{equation}
\mathbf{B}_{k} = X + j Y
\label{eq:Bk}
\end{equation}

The complex values of each frequency bin are then used to compute the magnitude of the corresponding bin:

\begin{equation}
|\mathbf{B}_k| = \sqrt{X^2 + Y^2}.
\label{eq:Bksqr}
\end{equation}

Thus, we construct four vectors for each sample: the raw in-phase ($\mathbf{I}$) and quadrature ($\mathbf{Q}$) time-domain vectors, and their corresponding FFT magnitude vectors. The time-domain in-phase sample vector can be represented as:

The time-domain in-phase vector for a single sample is represented as

\begin{equation}
\mathbf{y}_n^I = [\,y^I(0),\,y^I(1),\,\dots,\,y^I(N-1)\,]^{\mathsf T} \in \mathbb{R}^{N}.
\label{eq:12singsampi}
\end{equation}

Similarly, the time-domain quadrature vector is denoted by

\begin{equation}
\mathbf{y}_n^Q = [\,y^Q(0),\,y^Q(1),\,\dots,\,y^Q(N-1)\,]^{\mathsf T} \in \mathbb{R}^{N}.
\label{eq:13singsampq}
\end{equation}

We then compute the FFT of the raw I and Q vectors to obtain the discrete frequency bins $\mathbf{B}_k$, and the magnitude of each bin is calculated using Eqn.~\ref{eq:Bksqr}. This forms the remaining two vectors corresponding to the frequency-domain representation of the in-phase and quadrature components.

%----------------------------------------------------------
\begin{equation}
\resizebox{0.75\columnwidth}{!}{$
\lvert \mathbf{B}_k^{FI} \rvert
=
\bigl[
\lvert B^{FI}(0)\rvert,\;
\lvert B^{FI}(1)\rvert,\;
\dots,\;
\lvert B^{FI}(K-1)\rvert
\bigr]^T
$}
\;\in\;\mathbb{R}^K
\label{eq:14}
\end{equation}
%----------------------------------------------------------
%----------------------------------------------------------
\begin{equation}
\resizebox{0.75\columnwidth}{!}{$
\lvert \mathbf{B}_k^{FQ} \rvert
=
\bigl[
\lvert B^{FQ}(0)\rvert,\;
\lvert B^{FQ}(1)\rvert,\;
\dots,\;
\lvert B^{FQ}(K-1)\rvert
\bigr]^T
$}
\;\in\;\mathbb{R}^K
\label{eq:15}
\end{equation}
%----------------------------------------------------------

The time-domain in-phase and quadrature signals, along with their corresponding FFT magnitude vectors, are illustrated in Fig.~\ref{fig:rawsample} and Fig.~\ref{fig:fftmagnitude}, respectively. The complete data preprocessing workflow for obtaining the four vectors is summarized in Fig.~\ref{fig:dataprocessing}. This preprocessed dataset is then used as input for automatic modulation classification with our proposed deep learning model, HFEMCNet.

\section{Proposed HFEMCNet AMC Model}

We propose a lightweight multi-channel network, termed the Hybrid Frequency-Enriched Multi-Channel Network (HFEMCNet), designed for efficient and accurate automatic modulation classification. While the multi-channel convolutional long short-term deep neural network (MCLDNN) \cite{source24} achieves high classification accuracy across benchmark datasets, its computational complexity remains a significant challenge. Motivated by the strong performance of multi-channel architectures, HFEMCNet adopts a streamlined design with three independent channels, each specialized for complementary aspects of the input signals. The first channel processes the raw in-phase ($\mathbf{I}$) sequence and its corresponding FFT magnitude vector, while the second channel handles the quadrature ($\mathbf{Q}$) sequence along with its FFT magnitudes. The third channel receives the combined I/Q sequences and their FFT magnitudes to jointly capture cross-component dependencies. Each channel consists of three convolutional layers to extract temporal and spatial features, followed by an LSTM layer to model long-term temporal dependencies. Gaussian dropout is applied to enhance generalization, and a fully connected layer performs the final classification. Fig.~\ref{fig:framework} illustrates the architecture of HFEMCNet.

\subsection{CNN Layers}

Three independent channels are constructed by fusing the raw data, represented as $I$ and $Q$ vectors in Eqn.~\ref{eq:12singsampi} and Eqn.~\ref{eq:13singsampq}, with their corresponding FFT magnitude vectors in Eqn.~\ref{eq:14} and Eqn.~\ref{eq:15}. 
The $I/Q$ vectors and their FFT magnitudes serve as feature maps of dimension $\mathbb{R}^{128 \times 1}$ and are used as input vectors for the three independent channels. Feature concatenation layers are employed to fuse these vectors. The $I$ and $Q$ channels can be formally expressed as:

\begin{equation}
\mathrm{I\textsubscript{channel}} \;=\; \ y_n^I  \oplus \mathbf\lvert{B}_k^{FI}\rvert\ \in \mathbb{R}^{128 \times 2}\;
\end{equation}

\begin{equation}
\mathrm{Q\textsubscript{channel}} \;=\; \ y_n^Q  \oplus \mathbf\lvert{B}_k^{FQ}\rvert\; \in \mathbb{R}^{128 \times 2}\;
\end{equation}

Separate 1D convolutional layers are applied to the $I$ and $Q$ channels to capture the transient behavior of the waveform, spectral signatures, and harmonics in both channels.

\begin{equation}
C_{1D} = \mathrm{ReLU}\bigl( \text{I/Q channel} \ast W + b \bigr)
\end{equation}

Where \( W \in \mathbb{R}^{8 \times 2 \times 50} \) and \( C_{1D} \in \mathbb{R}^{128 \times 50} \) correspond to the I and Q channels, respectively. The input feature map vectors are reshaped to \( \mathbb{R}^{1 \times 128 \times 1} \) prior to forming a combined
channel in order to satisfy the requirements of 2D convolution. Three concatenation layers are employed to integrate the I/Q channels with their corresponding FFT magnitude feature maps:

\begin{equation}
\mathrm{Comb\textsubscript{IQ}} \;=\; \ y_n^I  \oplus \ y_n^Q \in \mathbb{R}^{2 \times 128\times 1}\;
\end{equation}
\begin{equation}
\mathrm{Comb\textsubscript{FFT}} \;=\; \mathbf\lvert{B}_k^{FQ}\rvert\ \oplus \mathbf\lvert{B}_k^{FI}\rvert\; \in \mathbb{R}^{2 \times 128\times 1}\;
\end{equation}
\begin{equation}
\mathrm{Comb\textsubscript{channel}} \;=\; Comb\textsubscript{IQ} \oplus Comb\textsubscript{FFT} \in \mathbb{R}^{4 \times 128\times 1}\;
\end{equation}
2D convolutional is applied to learn joint time/frequency patterns on combined channel.

\begin{equation}
C_{2D} = \mathrm{ReLU}\bigl( \text{Combined channel} \ast W_c + b_c \bigr)
\end{equation}

Where \( C_{2D} \) is obtained using convolutional kernels \( W_c \in \mathbb{R}^{2 \times 8 \times 1 \times 50} \), resulting in an output feature map of dimension \( C_{2D} \in \mathbb{R}^{4 \times 128 \times 50} \) after the 2D convolution. To reduce computational complexity and to make the proposed framework compact and lightweight, downsampling is performed after the Conv2D operation. The application of max-pooling on the combined channel reduces the overall computational complexity and number of parameters in the proposed framework. 

\begin{figure}[t]
    \centering
    \includegraphics[width=0.95\linewidth]{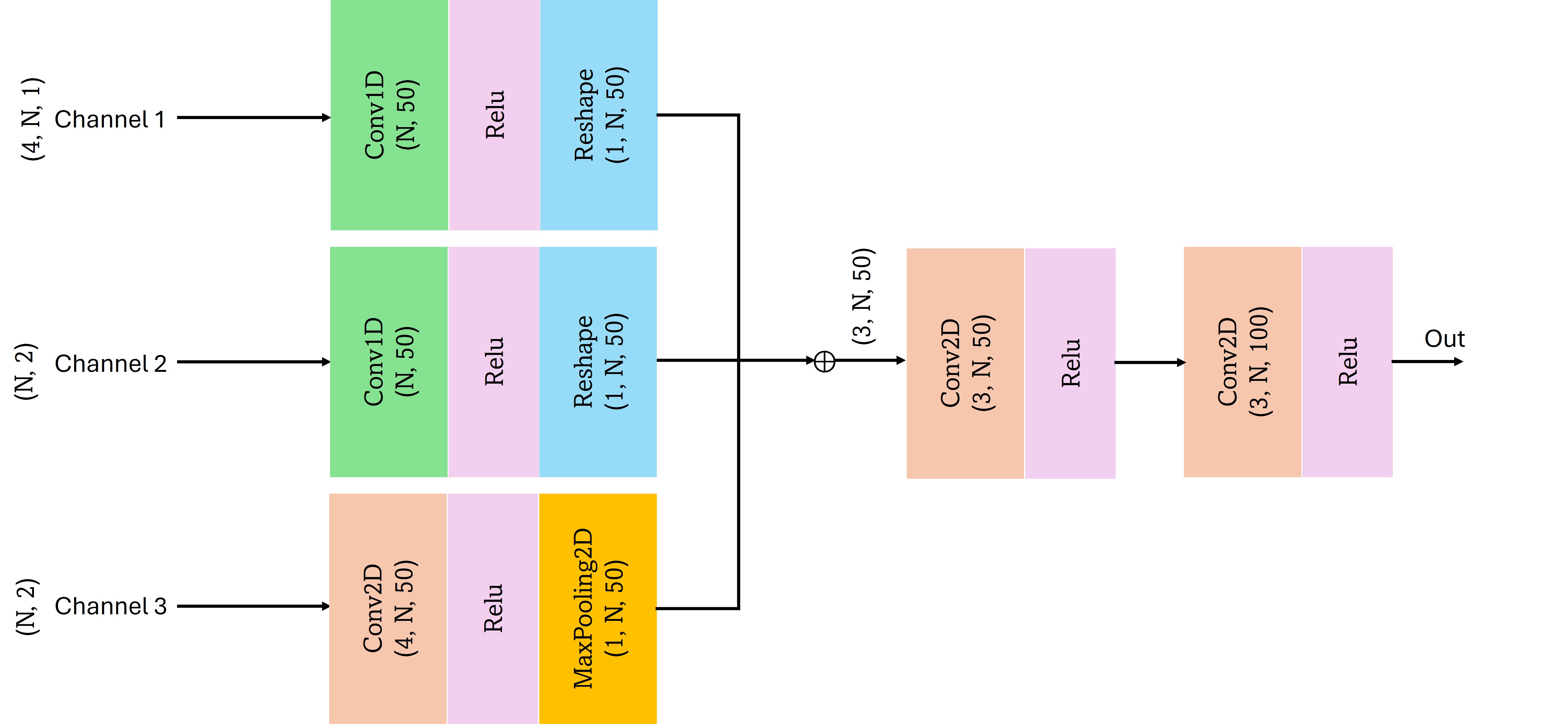}
    \caption{CNN Operation of HFEMCNet. The features extracted from each channel are fused via concatenation after the respective CNN layers. The fused representation is then refined by two stacked convolutional layers to produce a more abstract feature representation before being passed to the subsequent stage of the model}
    \label{fig:cnnoperations}
\end{figure}

The features extracted from each channel after passing through the respective CNN layers are fused again using concatenation. Subsequently, two stacked convolutional layers refine the previously extracted features and generate a more abstract representation before feeding them to the next stage of the model. The CNN layer operations are illustrated in Fig.~\ref{fig:cnnoperations}.

%----------------------------------------------------------------------------
%----------------------------------------------------------------------------

\subsection{LSTM and Classification}
Two LSTM layers are employed to extract long-term dependencies. LSTMs mitigate the vanishing gradient problem encountered in conventional RNNs by using a gated framework. Three gates input, forget, and output, regulate the flow of information, while the cell state serves as memory, carrying information across time steps. The first LSTM layer extracts long-term temporal dependencies, whereas the second LSTM layer collapses the sequence into a single vector for subsequent classification.
Post CNN layers, the 3D tensor is reshaped into a 2D sequence with shape \( (T, F) \), where \( T = 248 \) represents the number of time steps and \( F = 100 \) denotes the number of features per step. The reshaped sequence is provided as input to the first LSTM layer to capture temporal dependencies in the features extracted by the CNN layers. The output of
the first LSTM layer is a hidden state sequence \( H^{(1)} \in \mathbb{R}^{T \times 128} \), where the full return sequence is preserved. The second LSTM layer outputs a final hidden state \( h^{(2)} \in \mathbb{R}^{128} \), which encodes detailed temporal information.

% ------------------------------------------------------------------------------------------
\renewcommand{\arraystretch}{1.5}       % optional: tweak row height
\begin{table*}[t]
\caption{Details of $03$ Benchmark Datasets RML2016.10a, RML2016.10b and RML2018.01a} %\aafaq{20th scheme in 2018 ? elaborate the caption}}
\begin{center}
\begin{tabular}{c|c|c|c|c|c}
\hline
\textbf{Dataset Name}&\textbf{Number of Modulations}&\textbf{Modulation Schemes}&\textbf{Format per sample}&\textbf{Total Samples}&\textbf{SNR Range in dB}\\ \hline \hline 
RML2016.10a&        11&  \parbox[t]{4cm}{PSK, BPSK, CPFSK, 
GFSK, PAM4, 16QAM,AM-DSB, AM-SSB, 64QAM, QPSK, WBFM}& 2x128&220000&-20:2:18\\
\hline
RML2016.10b&   10&  \parbox[t]{4cm}{8PSK, BPSK, CPFSK, GFSK, PAM4, AM-DSB,
16QAM, 64QAM, QPSK, WBFM}& 2x128&1200000&-20:2:18\\ \hline
 RML2018.01a& 24& \parbox[t]{4cm}{OOK, 4ASK, 8ASK, BPSK, QPSK, 8PSK,
16PSK, 32PSK, 16APSK, 32APSK, 64APSK, 128APSK,
16QAM, 32QAM, 64QAM, 128QAM, 256QAM,
AM-SSB-WC, AM-SSB-SC, AM-DSB-WC, AM-DSB-SC,
FM, GMASK, OQPSK}& 2x1024&2555904&-20:2:30\\\hline
 \end{tabular}
\label{tabledatasets}
\end{center}
\end{table*}
% ------------------------------------------------------------------------------------------

A fully connected (dense) layer with 128 units and SELU activation is used, providing \( f \in \mathbb{R}^{128} \), which acts as a nonlinear transformation of the integrated spatiotemporal feature vector \( h^{(2)} \). The spatiotemporal feature maps extracted by the earlier CNN and LSTM layers provide a rich and robust representation for downstream classification. To mitigate overfitting, a dual dropout regularization mechanism is applied, combining standard dropout and Gaussian dropout. The dropout rate for both regularizations is set to 0.5. Standard dropout randomly deactivates neurons, whereas Gaussian dropout adds noise to the activations.

To generalize the proposed model ability a gaussian dropout is more effective instead of using only standard dropout. Gaussian dropout result smooth and continuous regularization instead of using the standard dropout (hard on/off). Use of zero mean gaussian noise enable to reduce bias in each forward pass. Accelerated convergence and improved training stability is achieved due continuous injection of noise. Moreover, sudden signal loss is avoided due nonzero activation.
For classification a DNN comprised of SELU and softmax activations are deployed. DNN layer computes the probability of each class \(\hat{y} \in \mathbb{R}^{C}\), where \(C\) representing the number of classes.

\begin{equation}
\hat{y} = \mathrm{Softmax}\bigl(W_{\mathrm{out}} \,.f + b_{\mathrm{out}}\bigr) \in \mathbb{R}^{C}
\end{equation}

\section{Experiments and Analysis}
\subsection{Datasets for Experimentation}

We evaluate our model extensively on three widely used benchmark datasets: RML2016a, RML2016b, and RML2018.01a. A summary of these datasets is also provided in Table~\ref{tabledatasets}.

%----------------------------------------------------------------------------
\begin{table*}[t]
\centering
\caption{Classification Accuracy vs SNR of 
Proposed HFEMCNet and State-of-the-Art DL Methods
 -- RML2016.10a and RML2016.10b Datasets} %\aafaq{it should be -20 db on both the tables ?}}
\resizebox{\textwidth}{!}{  % This resizes the table to fit the width of the page
\begin{tabular}{c|c|c|c|c|c|c|c|c|c|c|c|c|c|c|c|c|c|c|c|c}
\hline
\textbf{DL Model} & \multicolumn{20}{c}{\textbf{Unique SNR Levels-RML2016.10a Dataset}} \\ \hline \hline
                & \ -20 & -18 & -16 & -14 & -12 & -10 & -8 & -6 & -4 & -2 & 0 & 2 & 4 & 6 & 8 & 10 & 12 & 14 & 16 & 18 \\ \hline 
1DCNN-PF \cite{source17}       & 9.09 & 9.41 & 9.36 & 10.32 & 11.82 & 15.45 & 25.55 & 43.59 & 61.86 & 73.73 & 80.73 & 84.68 & 87.82 & 86.27 & 87.45 & 88.64 & 87.68 & 87.55 & 87.14 & 87.86 \\
 \hline
CGDNet \cite{source22}      & 9.32 & 8.95 & 9.55 & 10.64 & 13.73 & 20.18 & 32.14 & 51.00 & 63.32 & 75.14 & 79.09 & 81.14 & 82.82 & 81.82 & 84.18 & 84.05 & 82.55 & 82.91 & 82.41 & 82.86 \\
 \hline
CLDNN \cite{source23}      & 9.27 & 9.36 & 9.45 & 11.09 & 13.64 & 19.55 & 30.91 & 46.14 & 58.73 & 68.32 & 73.45 & 73.91 & 75.32 & 73.59 & 75.00 & 76.73 & 75.55 & 75.59 & 75.36 & 75.77 \\
 \hline
CNN1 \cite{source2}       & 9.00 & 9.41 & 10.00 & 12.18 & 17.09 & 25.32 & 37.68 & 54.36 & 64.82 & 74.41 & 78.14 & 79.59 & 81.86 & 79.68 & 81.59 & 82.23 & 81.18 & 80.36 & 79.82 & 81.64 \\
 \hline
CNN2 \cite{source13}       & 9.36 & 9.18 & 9.50 & 11.50 & 14.73 & 21.55 & 34.91 & 57.27 & 69.27 & 77.45 & 81.50 & 82.32 & 84.18 & 82.00 & 82.95 & 84.27 & 83.27 & 82.82 & 83.18 & 83.73 \\
 \hline
DAE \cite{source20}       & 9.05 & 9.18 & 10.05 & \textbf{12.77} & 16.77 & 24.82 & 37.36 & 50.86 & 60.41 & 71.45 & 79.09 & 81.64 & 84.95 & 84.86 & 85.32 & 85.45 & 86.00 & 84.18 & 83.86 & 85.18 \\
 \hline
DenseNet \cite{source14}       & 9.27 & 8.86 & 9.64 & 10.32 & 11.64 & 19.41 & 33.23 & 52.05 & 61.73 & 71.55 & 78.77 & 81.55 & 83.68 & 82.50 & 84.00 & 83.32 & 83.55 & 82.68 & 83.77 & 83.82 \\
 \hline
GRU \cite{source18}       & \textbf{10} & 9.50 & 9.90 & 12.00 & 15.00 & 24.00 & 31.00 & 54.00 & 59.00 & 73.00 & 83.00 & 86.00 & 87.00 & 86.00 & 87.00 & 88.00 & 86.00 & 86.50 & 86.50 & 86.50 \\
 \hline
ICAMC-Net \cite{source15}      & 9.14 & 9.23 & 9.14 & 9.64 & 13.09 & 20.09 & 31.82 & 51.50 & 63.77 & 73.68 & 80.64 & 81.86 & 84.64 & 84.45 & 85.05 & 85.95 & 85.36 & 84.59 & 83.59 & 84.91 \\
 \hline
LSTM \cite{source19}       & \textbf{10} & \textbf{9.82} & \textbf{10.68} & 13.41 & 16.50 & 22.55 & 35.32 & 51.73 & 62.95 & 75.86 & 85.45 & 88.45 & 91.23 & 90.64 & 91.41 & 91.09 & 90.82 & 90.55 & 90.73 & 91.05 \\
 \hline
MCLDNN \cite{source24}       & 9.18 & 9.27 & 9.82 & 12.18 & \textbf{18.23} & \textbf{25.41} & 40.73 & 54.45 & 65.91 & 79.95 & 88.36 & 89.95 & 91.95 & 90.77 & 92.41 & 91.59 & 91.77 & 90.95 & 91.00 & 91.77 \\
 \hline
MCNET \cite{source16}      & 9.32 & 9.32 & 9.82 & 11.18 & 16.23 & 24.36 & \textbf{38.86} & 52.73 & 62.41 & 72.36 & 77.95 & 79.82 & 82.50 & 81.59 & 82.00 & 84.95 & 83.64 & 82.09 & 81.68 & 83.09 \\
 \hline
PET-CGDNN \cite{source21}       & 9.68 & 9.55 & 9.91 & 11.95 & 15.86 & 24.91 & 38.77 & 53.73 & 66.00 & 76.18 & 84.55 & 88.05 & 90.18 & 88.77 & 89.50 & 90.23 & 89.95 & 90.09 & 89.45 & 90.23 \\
 \hline

 HFEMCNet (Ours)      & 9.32 & 9.46 & 9.50 & 12.09 & 15.91 & 24.77 & 37.36 & \textbf{54.95} & \textbf{70.14} & \textbf{82.00} & \textbf{90.18} & \textbf{91.59} & \textbf{92.59} & \textbf{92.95} & \textbf{93.46} & \textbf{93.00} & \textbf{94.14} & \textbf{93.55} & \textbf{92.59} & \textbf{92.68} \\

 \hline

\hline
\textbf{DL Model} & \multicolumn{20}{c}{\textbf{Unique SNR Levels-RML2016.10b Dataset}} \\ \hline
                & \ -20 & -18 & -16 & -14 & -12 & -10 & -8 & -6 & -4 & -2 & 0 & 2 & 4 & 6 & 8 & 10 & 12 & 14 & 16 & 18 \\ \hline \hline
1DCNN-PF \cite{source17}       & 10.52 & 10.54 & 11.64 & 12.75 & 14.39 & 20.32 & 30.50 & 47.41 & 64.82 & 76.90 & 84.73 & 88.41 & 89.73 & 90.32 & 90.54 & 90.86 & 91.31 & 91.13 & 90.97 & 91.21 \\

 \hline
CGDNet \cite{source22}      & 11.01 & 11.20 & 12.13 & 14.41 & 20.33 & 29.91 & 37.89 & 50.22 & 66.32 & 79.53 & 85.73 & 88.25 & 89.22 & 89.91 & 89.73 & 89.48 & 89.84 & 89.98 & 89.63 & 89.76 \\

 \hline
CLDNN \cite{source23}      & \textbf{11.13} & 10.99 & 12.23 & 14.03 & 19.13 & 28.93 & 39.99 & 53.34 & 67.73 & 77.15 & 80.78 & 83.03 & 84.00 & 84.02 & 83.97 & 84.28 & 84.15 & 84.24 & 83.72 & 84.02 \\

 \hline
CNN1 \cite{source2}       & 10.88 & 10.37 & 11.68 & 13.08 & 17.68 & 27.07 & 40.61 & 55.56 & 68.72 & 78.76 & 82.64 & 83.48 & 84.53 & 84.52 & 84.40 & 84.50 & 84.48 & 84.71 & 84.38 & 84.41 \\

 \hline
CNN2 \cite{source13}      & 10.93 & 10.93 & 12.20 & 14.58 & 18.89 & 26.50 & 37.78 & 53.73 & 70.00 & 79.99 & 83.22 & 84.22 & 84.97 & 85.58 & 85.50 & 85.47 & 85.51 & 85.44 & 85.19 & 85.23 \\

 \hline
DAE \cite{source20}       & 11.15 & 10.78 & 12.61 & \textbf{15.43} & \textbf{22.31} & 30.62 & 40.68 & 55.39 & 69.90 & 82.15 & 89.23 & 91.69 & 92.77 & 92.94 & 93.17 & 93.19 & 93.36 & 93.46 & 93.38 & 93.15 \\

 \hline
DenseNet \cite{source14}       & 10.51 & 10.44 & 11.23 & 12.34 & 14.87 & 24.10 & 38.65 & 53.88 & 64.88 & 73.08 & 81.75 & 86.02 & 88.48 & 89.13 & 89.79 & 89.13 & 89.15 & 89.67 & 89.34 & 89.67 \\

 \hline
GRU \cite{source18}       & 10.62 & 11.04 & \textbf{13.04} & 15.20 & 21.68 & 31.03 & 41.98 & 56.03 & 71.38 & 84.99 & 90.33 & 92.21 & 92.85 & 93.39 & 93.35 & 93.27 & 93.49 & 93.38 & 93.25 & 93.42 \\

 \hline
ICAMC-Net \cite{source15}      & 10.27 & 10.60 & 10.74 & 11.94 & 17.13 & 24.15 & 38.37 & 52.94 & 67.71 & 80.37 & 88.18 & 90.71 & 91.63 & 92.08 & 92.19 & 92.22 & 92.22 & 92.42 & 92.38 & 92.30 \\

 \hline
LSTM \cite{source19}       & 10.77 & 10.80 & 12.13 & 15.00 & 20.50 & \textbf{32.08} & \textbf{43.53} & 55.50 & 70.32 & 83.38 & 90.28 & 92.29 & 93.03 & 93.38 & 93.52 & 93.24 & 93.59 & 93.65 & 93.42 & 93.43 \\

 \hline
MCLDNN \cite{source24}       & 11.08 & \textbf{11.27} & 11.91 & 14.53 & 21.59 & 31.41 & 45.31 & \textbf{57.81} & 70.46 & 83.92 & 90.64 & 92.44 & 93.23 & 93.33 & 93.50 & 93.33 & 93.53 & 93.41 & \textbf{93.52} & 93.49 \\

 \hline
MCNET \cite{source16}      & 10.80 & 10.98 & 12.17 & 15.19 & 21.93 & 28.38 & 36.55 & 52.76 & 66.90 & 78.56 & 84.77 & 87.02 & 88.19 & 88.84 & 88.58 & 88.33 & 88.54 & 88.69 & 88.65 & 88.53 \\

 \hline
PET-CGDNN \cite{source21}     & 11.03 & 10.64 & 12.53 & 15.39 & 20.96 & 30.44 & 41.56 & 54.73 & 69.19 & 82.65 & 89.57 & 91.97 & 92.89 & 93.29 & 93.03 & 93.08 & 93.39 & 93.38 & 93.41 & 93.37 \\

 \hline
HFEMCNet (Ours)      & 10.77 & 10.85 & 12.57 & 15.22 & 21.70 & 30.93 & 42.83 & 56.41 & \textbf{72.94} & \textbf{85.84} & \textbf{91.20} & \textbf{92.61} & \textbf{93.51} & \textbf{93.45} & \textbf{93.57} & \textbf{93.47} & \textbf{93.66} & \textbf{94.05} & 93.51 & \textbf{93.91} \\
 \hline 
\end{tabular}
\label{tab:results2016datasets}
}

\end{table*}
%----------------------------------------------------------------------------

\subsubsection{RML2016}\label{rml2016}
Different versions of the RML2016 datasets are widely used for evaluating automatic modulation classification (AMC) models. The RML2016 dataset is generated using GNU Radio~\cite{source30}. 
Digital signals are synthesized using text from Gutenberg's \textit{Shakespeare} works, while audio sources are used for analog modulations. Channel modeling simulates real-world scenarios, including additive white Gaussian noise (AWGN), multipath effects, and fading. The signal-to-noise ratio (SNR) ranges from $-20$~dB to $18$~dB in steps of $2$~dB. Each sample contains $128$ discrete $I$ and $128$ discrete $Q$ values. RML2016.10a contains $11$ modulation schemes ($8$ digital and $3$ analog), whereas RML2016.10b contains $10$ modulation schemes ($8$ digital and $2$ analog). Each modulation scheme has $1000$ samples per SNR level in RML2016.10a. We adopt a 6:2:2 split for training, validation, and testing. Accordingly, $72000$ samples are randomly selected for training and $24000$ samples for validation and testing in RML2016.10a. Similarly, for RML2016.10b, $720000$ samples are used for training and $240000$ for validation and testing.

\subsubsection{RML2018.01a}\label{rml2018}
RML2018.01a is a comprehensive dataset comprising $24$ modulation schemes and over $2.5$ million examples, with $1024$ discrete $I/Q$ values per sample. The $24$ modulation schemes include $19$ digital and $5$ analog modulations, along with higher-order schemes such as $256-QAM$ and $256-APSK$. The dataset incorporates both synthetic channel impairments and over-the-air (OTA) recordings~\cite{source11} to emulate real-world conditions. SNR values range from $-20$~dB to $30$~dB in steps of $2$~dB, providing a broader SNR distribution compared with RML2016. We use the same 6:2:2 split for training, validation, and testing for our HFEMCNet deep learning AMC framework.

\subsection{Hyperparameter and Experimental Setup}

Our proposed model is trained for $10000$ epochs to ensure extensive learning. An early stopping callback with a patience of $50$ epochs is employed; training terminates if the validation loss does not improve within this window. The initial learning rate is set to $0.001$ and optimized using the Adam optimizer. If the validation loss does not improve for $5$ consecutive epochs, the learning rate is reduced by a factor of $1/2$. A batch size of $400$ is used. Both standard and Gaussian dropout layers are applied with a dropout rate of $0.5$. All experiments are conducted on an NVIDIA GeForce RTX 4090 GPU using TensorFlow~2.18.0 and standalone Keras~3.8.0.

\subsection{Benchmarking with State-of-the-Art AMC Models}
An extensive performance comparison with contemporary state-of-the-art (SoTA) deep learning-based AMC models is conducted to demonstrate the efficacy of our proposed HFEMCNet model. The models include  1DCNN-PF \cite{source17}, CGDNet \cite{source22}, CLDNN \cite{source23}, CNN1 \cite{source2}, CNN2 \cite{source13}, DAE \cite{source20}, DenseNet \cite{source14}, GRU \cite{source18}, IC-AMCNet \cite{source15}, LSTM \cite{source19}, MCLDNN \cite{source24},  MCNET \cite{source16} and PET-CGDNN \cite{source21}.

% \subsubsection{Classification Accuracy}
The modulation classification performance of the proposed HFEMCNet framework is evaluated on three benchmark datasets: RML2016.10a, RML2016.10b, and RML2018.01a to assess its generalization across a wide dynamic range of radio signals. Table~\ref{tab:results2016datasets} reports the classification accuracy of the proposed HFEMCNet and the aforementioned state-of-the-art DL AMC models at each SNR level for the RML2016.10a and RML2016.10b datasets. As evident from Table~\ref{tab:results2016datasets}, the proposed HFEMCNet outperforms all contemporary SOTA methods by significant margins across most SNR levels and performs comparably at the remaining levels, demonstrating the effectiveness of the proposed framework.

%----------------------------------------------------------------------------
\begin{table*}[ht]
%\footnotesize
\centering
\caption{Classification Accuracy vs SNR of 
Proposed HFEMCNet and State-of-the-Art DL Methods - RML2018.01a Dataset} %\aafaq{try landscape full page table and see if it loos better ?}}
%\resizebox{\textwidth}{!}{  % This resizes the table to fit the width of the page
\begin{tabular}{c|c|c|c|c|c|c|c|c|c|c|c|c|c}
\hline
\textbf{DL Model} & \multicolumn{13}{c}{\textbf{Unique SNR Levels-RML2018.01a Dataset}} \\ \hline
                & \ -20 & -18 & -16 & -14 & -12 & -10 & -8 & -6 & -4 & -2 & 0 & 2 & 4 \\ \hline \hline
1DCNN-PF \cite{source17}      & 4.15 & 4.24 & 4.12 & 4.42 & 5.05 & 6.22 & 9.57 & 15.27 & 23.16 & 33.91 & 40.73 & 48.46 & 53.82 \\
 \hline
CGDNet \cite{source22}     & 4.18 & 4.43 & 4.96 & 6.25 & 9.42 & 13.96 & 17.61 & 23.26 & 29.41 & 37.43 & 46.65 & 56.65 & 66.63 \\
 \hline
CLDNN \cite{source23}       & 4.40 & 4.54 & 5.05 & 6.54 & 8.85 & 13.45 & 17.74 & 22.61 & 25.79 & 30.48 & 38.31 & 41.66 & 47.13 \\

 \hline
CNN1 \cite{source2}        & 4.52 & 4.46 & 5.09 & 6.53 & 9.83 & 13.44 & 16.03 & 21.65 & 29.50 & 37.57 & 44.62 & 51.73 & 57.61 \\

 \hline
CNN2 \cite{source13}      & 4.35 & 4.82 & 5.35 & \textbf{7.26} & 10.38 & 13.72 & 17.97 & 24.23 & 31.47 & 41.12 & 49.64 & 58.21 & 66.50 \\
 
 \hline

DenseNet \cite{source14}       & 4.48 & 4.33 & 4.76 & 4.43 & 5.45 & 9.96 & 14.18 & 20.93 & 28.03 & 38.62 & 50.82 & 57.23 & 62.93 \\

 \hline
GRU \cite{source18}      & 4.36 & 4.60 & 4.78 & 6.12 & 7.86 & 11.40 & 16.77 & 26.45 & 23.70 & 44.83 & 57.34 & 69.12 & 82.35 \\

 \hline
ICAMC-Net \cite{source15}     & 4.41 & 4.38 & 4.43 & 5.39 & 6.97 & 10.58 & 14.52 & 22.14 & 29.84 & 39.09 & 51.61 & 61.35 & 73.18 \\

 \hline
LSTM \cite{source19}        & 4.25 & 4.66 & 5.27 & 6.85 & 9.88 & 13.96 & 20.01 & 27.17 & 34.67 & 46.82 & 59.39 & 71.29 & 82.82 \\

 \hline
MCLDNN \cite{source24}       & 4.20 & 4.35 & 5.10 & 5.78 & 8.44 & 12.07 & 18.71 & 26.40 & 34.42 & 45.80 & 58.48 & 70.20 & 82.32 \\

 \hline
MCNET \cite{source16}       & 4.33 & 4.31 & 4.79 & 5.96 & 8.04 & 11.40 & 16.97 & 25.39 & 32.54 & 41.13 & 52.82 & 61.34 & 70.81 \\

 \hline
PET-CGDNN \cite{source21}        & 4.65  & 4.75  & 5.42  & 6.93  & 8.55  & 13.16  & 18.08  & 24.89  & 33.90  & 45.21  & 58.13  & 69.59  & 81.12 \\

 \hline

 HFEMCNet (Ours) & \textbf{5.01} & \textbf{5.32} & \textbf{5.92} & 7.23 & \textbf{11.12} & \textbf{15.21} & \textbf{20.56} & \textbf{27.99} & \textbf{37.62} & \textbf{49.27} & \textbf{60.12} & \textbf{71.76} & \textbf{83.03} \\

 \hline

\hline
\textbf{DL Model} & \multicolumn{13}{c}{\textbf{Unique SNR Levels-RML2018.01 Dataset}} \\ \hline \hline
                & \ 6 & 8 & 10 & 12 & 14 & 16 & 18 & 20 & 22 & 24 & 26 & 28 & 30 \\ \hline
1DCNN-PF \cite{source17}      & 60.03 & 67.35 & 70.60 & 72.67 & 73.46 & 73.34 & 74.04 & 73.96 & 73.62 & 74.14 & 74.24 & 73.85 & 73.73 \\
 \hline
CGDNet \cite{source22}      & 76.56 & 83.15 & 85.90 & 87.14 & 87.09 & 87.49 & 87.35 & 87.54 & 87.26 & 87.18 & 87.43 & 87.62 & 87.33 \\

 \hline
CLDNN \cite{source23}       & 55.93 & 58.17 & 60.59 & 61.62 & 61.87 & 60.14 & 60.04 & 60.70 & 60.33 & 60.87 & 60.36 & 60.62 & 61.39 \\

 \hline
CNN1 \cite{source2}        & 62.78 & 65.70 & 67.02 & 68.01 & 67.83 & 68.53 & 68.70 & 68.14 & 67.87 & 68.21 & 68.46 & 68.20 & 67.49 \\

 \hline
CNN2 \cite{source13}      & 71.22 & 76.90 & 81.28 & 81.85 & 81.85 & 82.31 & 82.67 & 83.26 & 82.59 & 83.06 & 82.59 & 82.37 & 82.94 \\
 
 \hline

DenseNet \cite{source14} & 70.61 & 79.71 & 85.58 & 86.01 & 87.91 & 90.49 & 91.49 & 91.93 & 90.34 & 91.23 & 90.37 & 90.86 & 90.19 \\

 \hline
GRU \cite{source18}      & 90.52 & 95.64 & 96.89 & 97.12 & 97.78 & 97.97 & 97.09 & 97.15 & 97.51 & 98.25 & 98.10 & 97.86 & 98.09 \\

 \hline
ICAMC-Net \cite{source15}      & 84.56 & 91.71 & 94.05 & 95.16 & 95.60 & 95.89 & 95.81 & 95.78 & 95.69 & 95.89 & 95.63 & 95.68 & 95.60 \\

 \hline
LSTM \cite{source19}        & 92.27 & \textbf{96.52} & 97.81 & 98.11 & 98.05 & 98.21 & 98.26 & 98.09 & 98.35 & 98.11 & 98.20 & 98.12 & 98.23 \\

 \hline
MCLDNN \cite{source24}       & 91.70 & 95.99 & 97.40 & 97.96 & 97.95 & 98.20 & 98.13 & 98.26 & 98.07 & 98.09 & 98.24 & 98.24 & 98.12 \\

 \hline
MCNET \cite{source16}       & 80.41 & 87.76 & 91.46 & 92.78 & 93.25 & 93.59 & 93.60 & 93.93 & 93.49 & 93.70 & 93.84 & 93.82 & 93.58 \\

 \hline
PET-CGDNN \cite{source21}      & 91.03  & 95.82  & 97.05  & 97.84  & 97.66  & 98.09  & 97.91  & 98.03  & 98.10  & 98.03  & 98.16  & 97.94  & 98.11 \\

 \hline

 HFEMCNet (Ours) & \textbf{92.58} & 96.49 & \textbf{98.16} & \textbf{98.65} & \textbf{98.64} & \textbf{98.73} & \textbf{98.74} & \textbf{98.62} & \textbf{98.89} & \textbf{98.87} & \textbf{98.79} & \textbf{98.80} & \textbf{98.74} \\
 \hline
\end{tabular}
\label{tab:results2018dataset}
% }
\end{table*}
%----------------------------------------------------------------------------

Similarly, the classification performance on the RML2018.01a dataset is summarized in Table~\ref{tab:results2018dataset}. The proposed HFEMCNet achieves exceptional results, outperforming contemporary state-of-the-art models across almost all the SNR ranges. Given the realistic, over-the-air nature of this dataset, these results further demonstrate the superiority and robustness of the proposed framework. 

Overall classification accuracy (average accuracy), computed as the mean across all SNR levels, provides a quantitative measure of the performance of contemporary state-of-the-art (SOTA) models, including our proposed framework, across three benchmark datasets. Our proposed model, HFEMCNet, consistently achieves the highest classification accuracy across all datasets. In contrast, some existing models exhibit significant performance degradation on large-scale, real-world datasets, such as RML2018.01.

Analysis of Tables~\ref{tab:results2016datasets}, \ref{tab:results2018dataset} and \ref{tab:resultsaverage} indicates that HFEMCNet outperforms competing frameworks over almost the entire SNR range. This observation demonstrates the robustness and generalization capability of our model under varying channel conditions. Table IV further highlights that deep learning based AMC  architectures such as LSTM~\cite{source19}, MCLDNN~\cite{source24}, and PETCGDNN~\cite{source21} also maintain competitive performance across all datasets, albeit slightly below HFEMCNet.

\begin{figure}[t]
    \centering
    \includegraphics[width=1\linewidth]{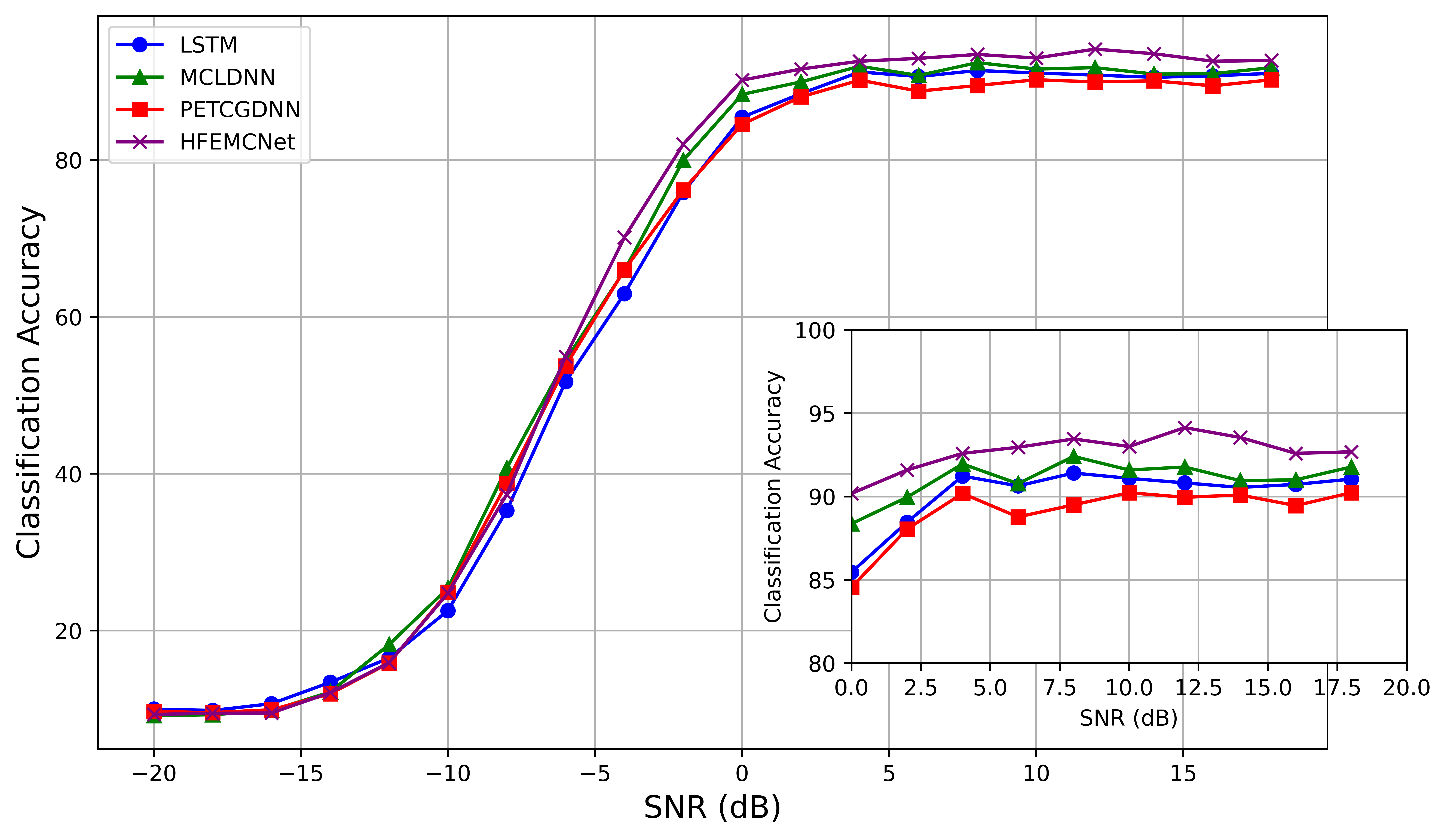}
    \caption{Classification Accuracy Vs SNR - RML2016.10a}
    \label{fig:classificationaccuracy}
\end{figure}

\begin{figure}[t]
    \centering
    \includegraphics[width=1\linewidth]{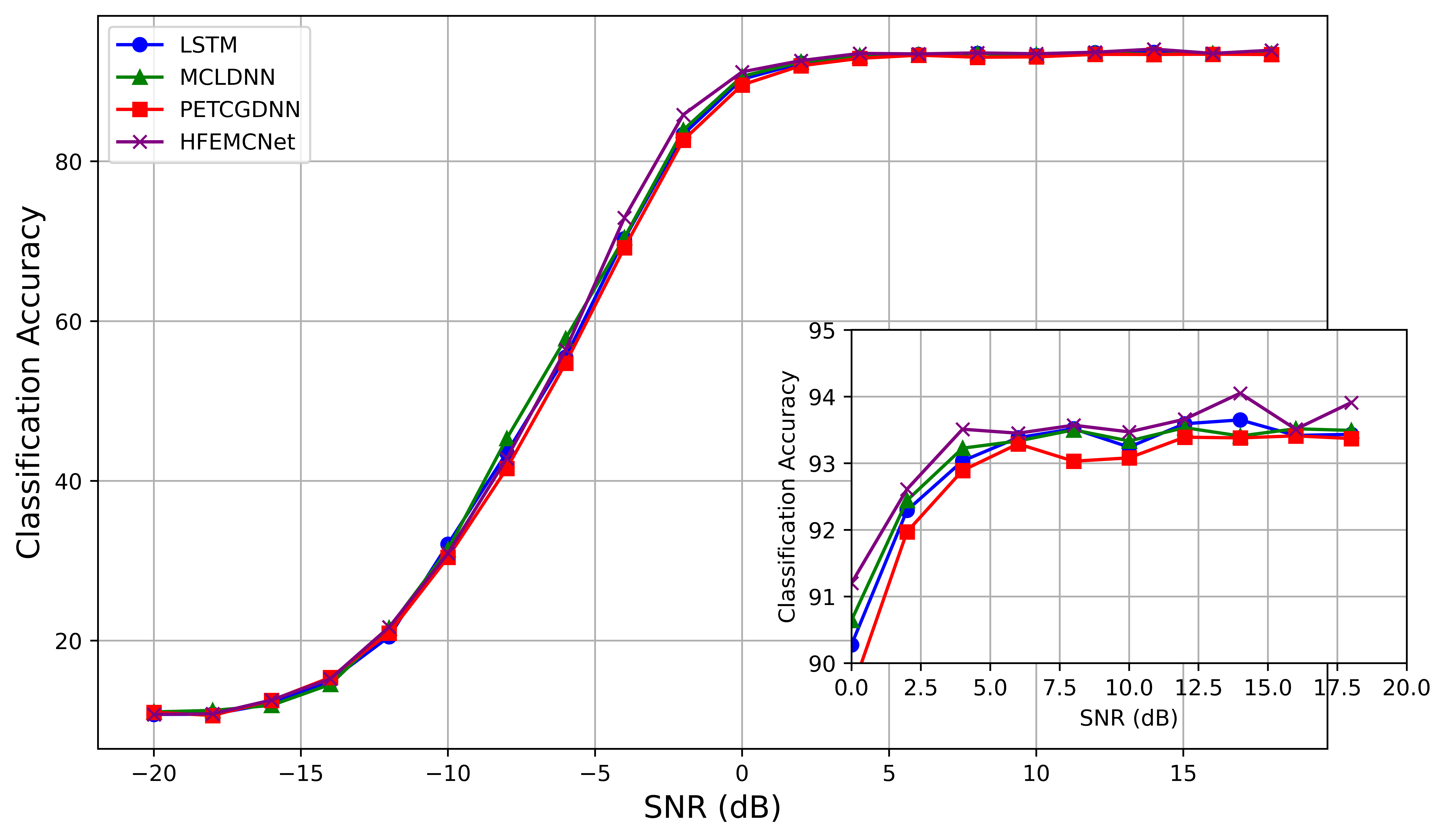}
    \caption{Classification Accuracy Vs SNR - RML2016.10b}
    \label{fig:classaccu201610b}
\end{figure}

\begin{figure}[t]
    \centering
    \includegraphics[width=1\linewidth]{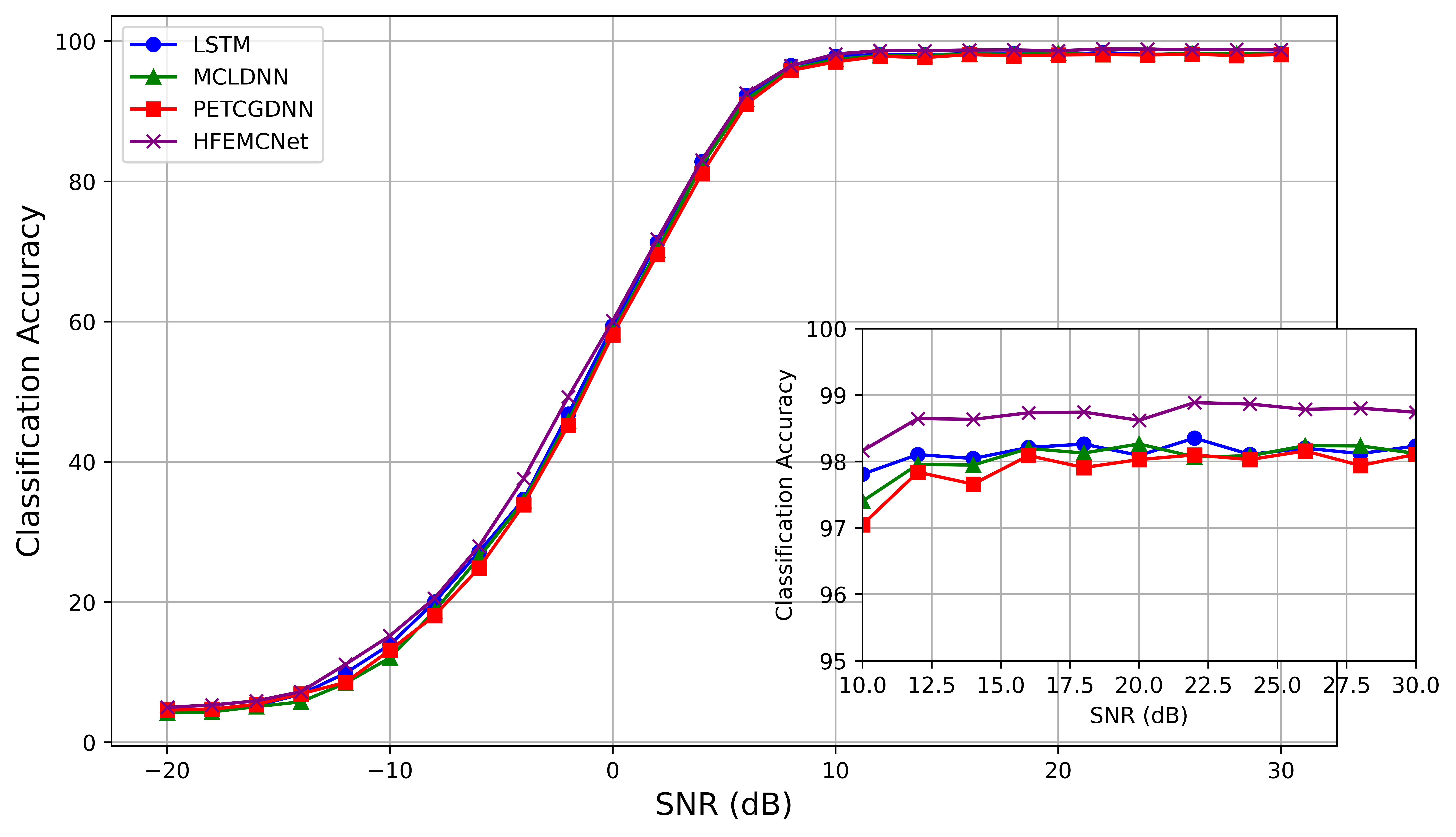}
    \caption{Classification Accuracy Vs SNR - RML2018.01a}
    \label{fig:classaccu201801a}
\end{figure}

Consequently, for better readability, only the classification performance curves of the top-performing models i.e., LSTM~\cite{source19}, MCLDNN~\cite{source24}, PETCGDNN~\cite{source21}, and HFEMCNet are shown in Figs.~\ref{fig:classificationaccuracy}, \ref{fig:classaccu201610b}, and \ref{fig:classaccu201801a} for RML2016.10a, RML2016.10b, and RML2018.01, respectively. As depicted by the figures, HFEMCNet’s superior performance and robustness, demonstrate consistently superior classification accuracy across the full SNR spectrum and all the datasets.

%----------------------------------------------------------
\begin{table*}[ht]
\centering
%\footnotesize
\caption{Comparison of Highest and Average Accuracies of State-of-the-Art DL Method with Proposed HFEMCNet} %\aafaq{maintain textwidth}}
\begin{tabular}{c|c|c|c}
\hline
\textbf{Model Name} & \textbf{Highest/ Average Accuracy} & \textbf{Highest/ Average Accuracy} & \textbf{Highest/ Average Accuracy} \\
& \textbf{RML2016.10a}& \textbf{RML2016.10b}& \textbf{RML2018.01a}\\
\hline \hline
1DCNN-PF \cite{source17} & 88.64 / 56.8 & 91.31 / 59.95 & 74.24 / 45.70 \\
\hline
CGDNet \cite{source22} & 84.18 / 55.89 & 89.98 / 61.22 & 87.62 / 55.38 \\
\hline
CLDNN \cite{source23} & 76.73 / 51.34 & 84.28 / 58.54 & 61.87 / 40.35 \\
\hline
CNN1 \cite{source2} & 82.23 / 56.02 & 84.71 / 58.82 & 68.70 / 45.39 \\
\hline
CNN2 \cite{source13} & 84.27 / 57.25 & 85.58 / 59.29 & 83.26 / 53.46 \\
\hline
DenseNet \cite{source14} & 84 / 55.77 & 89.79 / 59.81 & 91.93 / 55.50 \\
\hline
GRU \cite{source18} & 88 / 58.00 & 93.49 / 64.29 & 98.25 / 48.50 \\
\hline
ICAMC-Net \cite{source15} & 85.95 / 56.61 & 92.42 / 62.03 & 95.89 / 62.30 \\
\hline
LSTM \cite{source19} & 91.41 / 60.51 & 93.65 / 64.19 & 98.35 / 63.67 \\
\hline
MCLDNN \cite{source24} & 92.41 / 61.78 & 92.41 / 64.49 & 98.26 / 63.18 \\
\hline
MCNET \cite{source16} & 84.95 / 56.30 & 88.84 / 60.72 & 93.93 / 59.03 \\
\hline
PET-CGDNN \cite{source21} & 90.23 / 60.38 & 93.41 / 63.82 & 98.16 / 63.01 \\
\hline
HFEMCNet (Ours) & \textbf{94.14~/ 62.61} & \textbf{94.05~/ 64.65} & \textbf{98.89~/ 64.42} \\
\hline
\end{tabular}
\label{tab:resultsaverage}
\end{table*}
%----------------------------------------------------------

\begin{figure}[t]
    \centering
    \includegraphics[width=.8\linewidth]{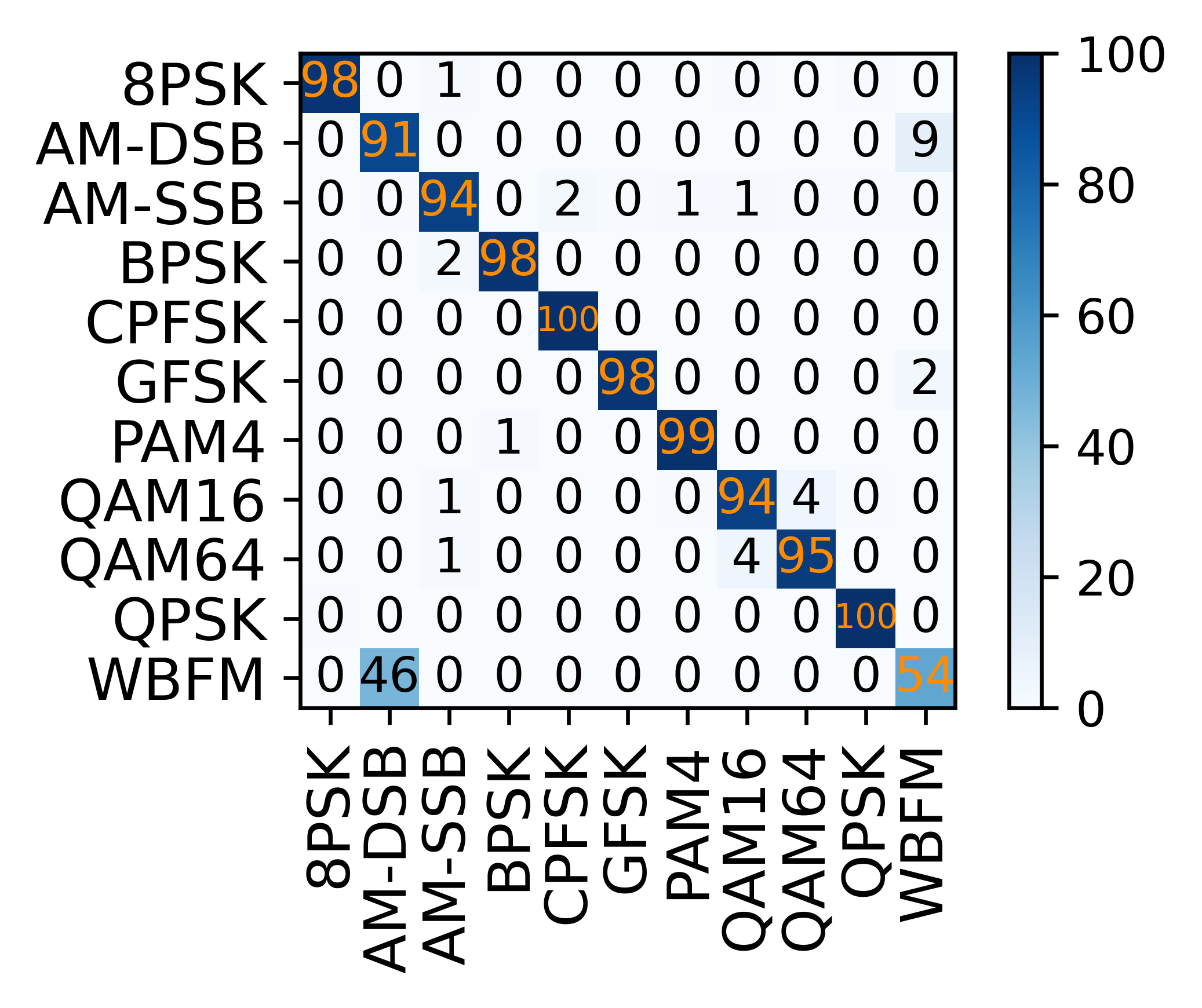}
    \caption{Confusion Matrix - RML2016.10a}
    \label{fig:confmatrix201610a}
\end{figure}

The confusion matrix is used to identify which modulation schemes are most challenging for deep learning based AMC and to gain deeper insight into model misclassifications under specific SNR conditions. In the matrix, actual labels are represented along the horizontal axis, and predicted labels along the vertical axis. A moderate SNR of 4 dB is selected to evaluate the model performance, providing a balanced scenario where both noise and signal features influence classification. For the RML2016.10a dataset, confusion occurs primarily between the analog modulation schemes AM-DSB and WBFM, likely due to their similar spectral characteristics, while all other modulation schemes exhibit high detection probabilities. Additionally, confusion between higher-order modulations, such as 16-QAM and 64-QAM, is notable, reflecting the inherent difficulty of discriminating closely spaced constellation points in DL-based AMC. The confusion matrix for RML2016.10a is shown in Fig.~\ref{fig:confmatrix201610a}.

\begin{figure}[t]
    \centering
    \includegraphics[width=.8\linewidth]{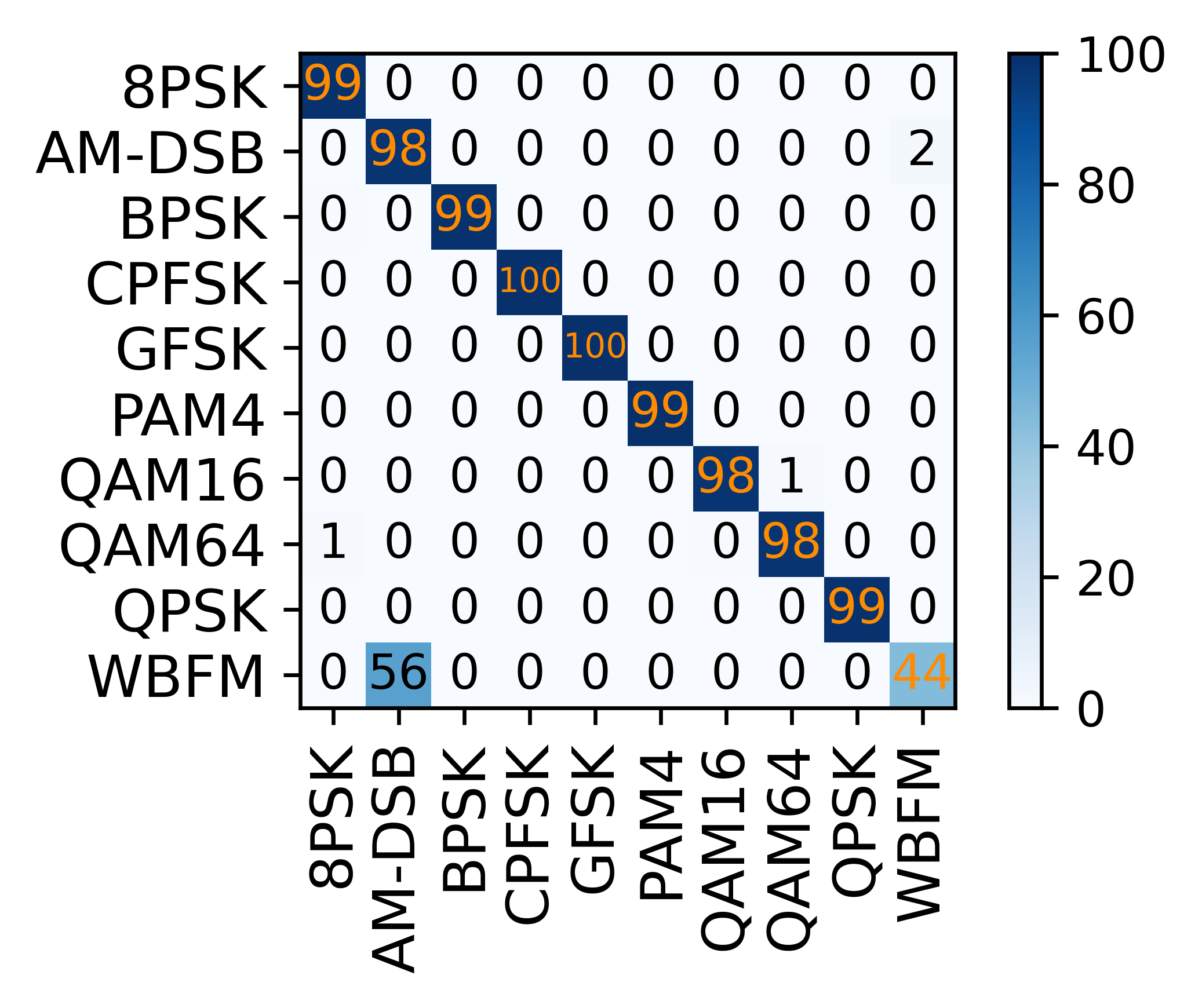}
    \caption{Confusion Matrix - RML2016.10b}
    \label{fig:confmatrix201610a2}
\end{figure}

As previously noted, the RML2016.10b dataset contains inputs similar to RML2016.10a, and its confusion matrix exhibits nearly identical patterns. The confusion between AM-DSB and WBFM persists. However, the overall classification performance is improved due to the larger number of training samples, which enables the model to better capture subtle modulation-specific features. The confusion matrix for RML2016.10b is presented in Fig.~\ref{fig:confmatrix201610a2}.

\begin{figure}[t]
    \centering
    \includegraphics[width=.8\linewidth]{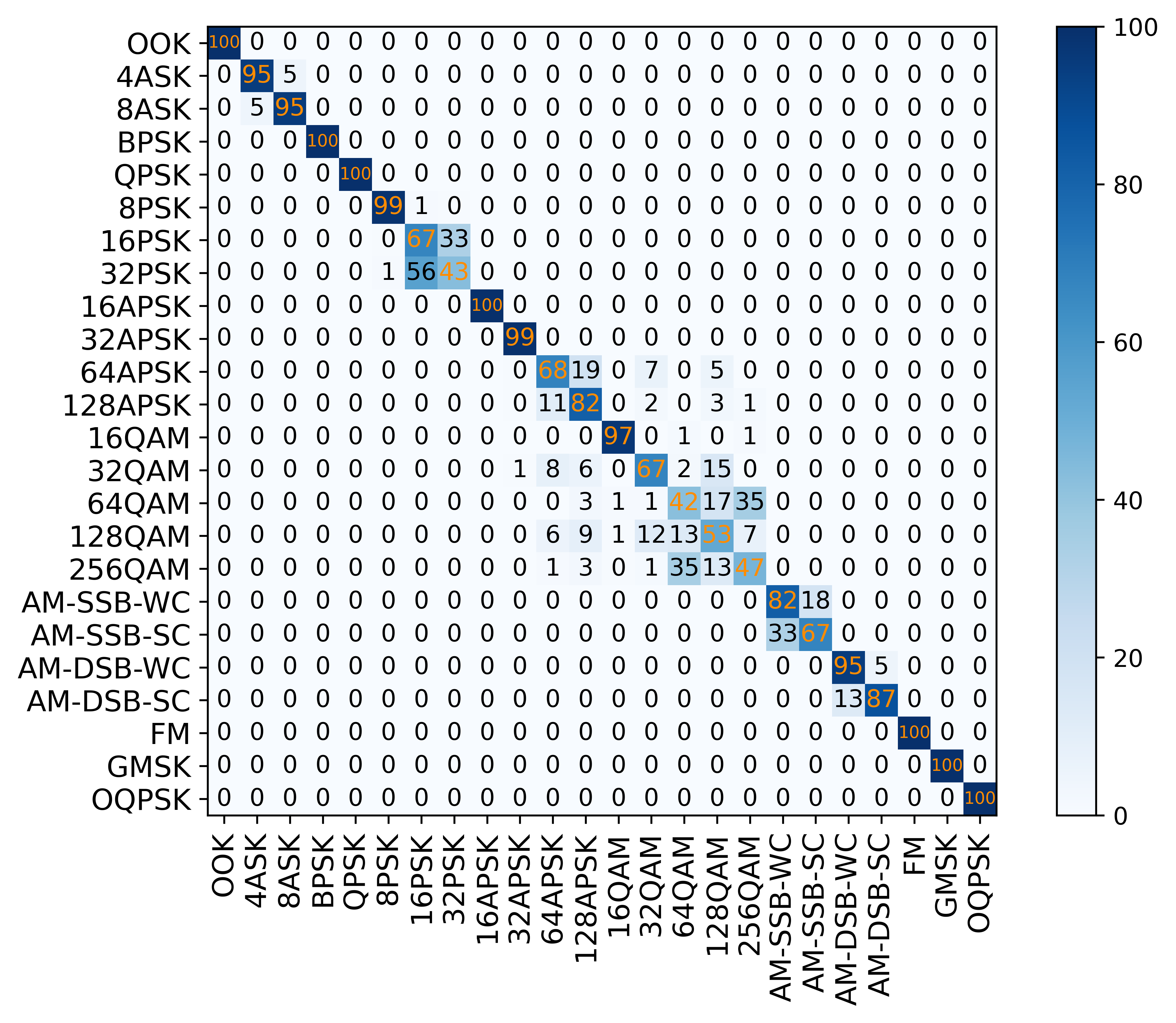}
    \caption{Confusion Matrix - RML2018.01a}
    \label{fig:confmatrix201801a}
\end{figure}

The RML2018.01a dataset contains a larger number of modulation schemes compared with the RML2016 datasets. The confusion matrix for RML2018.01a is shown in Fig.~\ref{fig:confmatrix201801a}. It reveals notable misclassifications among different variants of QAM and PSK modulations, as well as confusion in analog modulation schemes. This behavior indicates that variants of the same modulation type are more susceptible to misclassification due to additive noise and channel impairments during propagation. However, as the SNR increases, the classification accuracy improves significantly, demonstrating the model’s robustness under high-quality signal conditions.

%----------------------------------------------------------------------------
%----------------------------------------------------------------------------
%----------------------------------------------------------------------------
%----------------------------------------------------------------------------

\subsection{Computational Complexity Analysis}

To assess the computational complexity and efficiency of the proposed HFEMCNet, we evaluate the model using standard complexity metrics. These metrics quantitatively characterize memory consumption, computational cost, and hardware resource utilization of the deep neural network architecture. HFEMCNet is compared with MCLDNN \cite{source24}, given their architectural similarity and comparable high performance across all datasets. The analysis is conducted on the RML2016.10a dataset, with results summarized in Table~\ref{tab:complexityanalysis}. Our proposed model demonstrates superior performance across most evaluated metrics, including model complexity, memory efficiency (weights, CPU, and GPU usage), and tail latency, indicating its potential for resource-constrained environments and real time applications. Detailed evaluation of each computational metric and comparative analysis between HFEMCNet and MCLDNN are presented below.

\subsubsection{Total Parameters}

The total number of trainable parameters reflects the model capacity, memory footprint, and indirectly influences inference efficiency. In comparison with the MCLDNN, HFEMCNet contains $339463$ parameters, which is significantly lower than MCLDNN ($405175$). This reduction indicates a more parameter-efficient architecture, resulting in lower storage overhead and improved deployability under resource-constrained environments.

\subsubsection{Inference Latency}

Median latency (50th percentile) reflects typical inference speed, critical for latency-sensitive applications, while tail latency (99th percentile) indicates worst-case performance, relevant for safety-critical systems. MCLDNN achieves a slightly lower median latency (32.27 ms) than HFEMCNet (34.68 ms), suggesting marginally faster typical inference. However, HFEMCNet exhibits significantly lower tail latency (42.97 ms) compared to MCLDNN (57.95 ms), demonstrating improved predictability and robustness under worst-case conditions. The reduced tail latency of HFEMCNet can be attributed to its more compact and parameter-efficient architecture, which limits variability in memory and computation overhead during inference.

%----------------------------------------------------------------------------
%----------------------------------------------------------------------------

%----------------------------------------------------------------------------
%----------------------------------------------------------------------------
%----------------------------------------------------------------------------
%----------------------------------------------------------------------------
%----------------------------------------------------------------------------
%----------------------------------------------------------------------------

\subsubsection{Throughput}

Throughput, measured in samples per second, reflects the overall inference efficiency and scalability under high-load conditions. While HFEMCNet is lighter and more parameter-efficient, it achieves slightly lower throughput ($1859.15$ samples/s) than MCLDNN ($2679.29$ samples/s). This difference arises from the architectural design of HFEMCNet, which prioritizes reduced tail latency and predictable memory access over maximal batch processing speed, ensuring better real-time and resource-constrained performance.

\subsubsection{Memory Footprint}

Peak GPU memory and host-side RAM (RSS) critically determine deployment feasibility on memory-constrained hardware. HFEMCNet requires $4577.80$ MB of GPU memory, significantly lower than MCLDNN ($6543.96$ MB), and consumes $2221.81$ MB of host RAM, compared to $2451.44$ MB for MCLDNN. The reduced memory footprint stems from HFEMCNet’s parameter-efficient design and compact intermediate feature maps, making it more suitable for real-time inference on resource-limited systems.

% -----------------------------------------------------------------------------------
\begin{table}[ht]
  \centering
  \caption{Computational Complexity Analysis Between MCLDNN and Proposed HFEMCNet} %\aafaq{maintain columnwidth}}
  \label{tab:comp_complexity}
  \begin{tabular}{c|c|c}
    \hline
    \textbf{Metric}                     & \textbf{HFEMCNet (Ours)} & \textbf{MCLDNN} \\ 
    \hline \hline
    Total parameters                   & 339463            & 405175          \\ 
    \hline
    Weight file size (MB)              & 3.99              & 4.73            \\ 
    \hline
    Weights memory (MB)                & 1.29              & 1.55            \\ 
    \hline
    Latency P50 (ms)                   & 34.68             & 32.27           \\ 
    \hline
    Latency P99 (ms)                   & 42.97             & 57.95           \\ 
    \hline
    Throughput (samples/s)             & 1859.15           & 2679.29         \\ 
    \hline
    Peak RSS memory (MB)               & 2221.81           & 2451.44         \\ 
    \hline
    GPU peak memory (MB)               & 4577.80           & 6543.96         \\ 
    \hline
  \end{tabular}
  \label{tab:complexityanalysis}
\end{table}
% -----------------------------------------------------------------------------------

\subsection{Ablation Study}
\begin{figure}[t]
    \centering
    \includegraphics[width=1\linewidth]{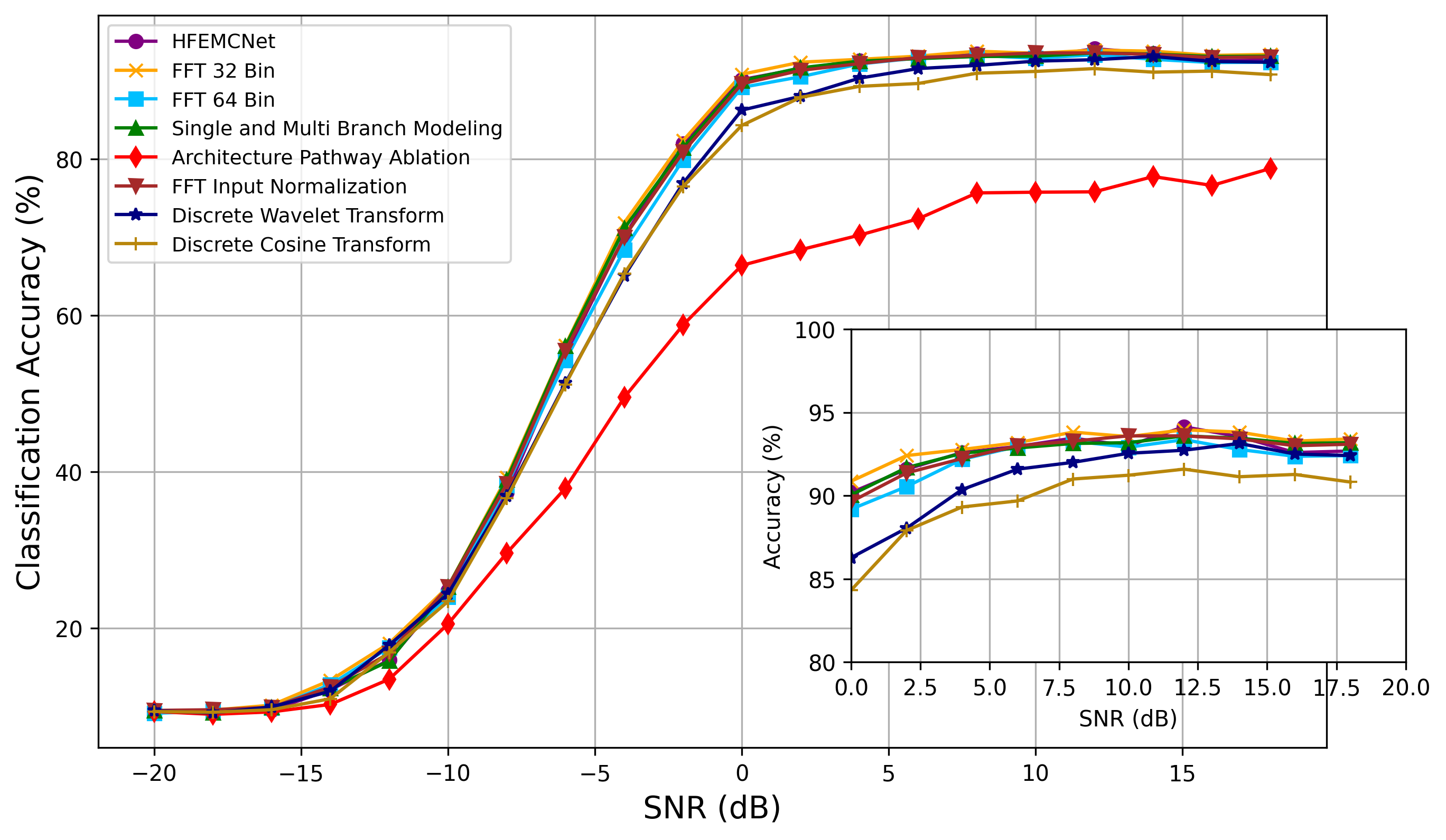}
    \caption{Results of Ablation Study}
    \label{fig:ablationstudyresults}
\end{figure}

We have conducted the extensive ablation experiments to ascertain the efficacy of FFT magnitude resolution reduction, single/ multi branch modeling, architecture pathway ablation, FFT input normalization and alternative frequency transforms such as Discrete Cosine Transform (DCT) and Discrete Wavelet Transform (DWT). Our proposed model achieved higher classification performance serving as a baseline for comparison. 
Reducing the FFT magnitude input resolution demonstrated minimal degradation, a marginal accuracy drop (62.564\%, $\Delta$ = -0.048\%) is observed when we have used 32 FFT bins. However, using 64 FFT bins resulted in a slightly more noticeable decline (61.446\%, $\Delta$ = -1.166\%), which has highlighted that model is moderate sensitivity to spectral resolution.
The Single/Multi-Branch Modeling ablation, in which we have only applied the FFT magnitudes to combined branch, while I and Q branches are using raw I and Q data only. This design reflects the effect of explicit branch configuration, which in slight decrease of classification performance (62.395\%, $\Delta$ = -0.216\%), reinforcing the model's robustness to modest structural adjustments. However, in architecture pathway ablation in which we have removed I and Q branches altogether, led to significant drop of classification accuracy (50.641\%, $\Delta$ = -11.970\%). Therefore, it is ascertained that all three branches are critical for optimal classification performance. 
The application of input normalization to FFT bins demonstrated a minor, but consistent, reduction in accuracy (62.041\%, $\Delta$ = -0.570\%). This can be attributed to the fact that input data is already normalized and further normalization of frequency bins is not necessary. We have also evaluated the alternative frequency-domain representations/ transforms namely Discrete Wavelet Transform (DWT) (59.529\%, $\Delta$ = -3.082\%) and Discrete Cosine Transform (DCT) (58.039\%, $\Delta$ = -4.573\%). Use of both the transforms have exhibited the considerable drop in classification accuracy compared to the FFT-based baseline, highlighting the superior discriminative capabilities of FFT-derived features for this task. 
Overall, these ablation results clearly emphasize the importance of maintaining explicit multi-pathway architectures and carefully selecting frequency-domain representations to achieve optimal performance in modulation classification tasks. Table~\ref{tab:ablationstudy} summarizes the results of ablation experiments and classification performance of each experiment is shown in Fig.~\ref{fig:ablationstudyresults}.

% ----------------------------------------------------------------------
\begin{table}[t]
\centering
\caption{Ablation Analysis}
\label{tab:ablation-accuracy}
\begin{tabular}{l|c|c}
\hline
\textbf{Ablation Name}                      & \textbf{Overall Accuracy} & \textbf{$\Delta$ Accuracy} \\
\hline \hline
HFEMCNet           & 62.611 & -- \\\hline
32 FFT Bins                       & 62.564 & -0.048 \\\hline
64 FFT Bins                      & 61.446 & -1.166 \\\hline
Single/Multi-Branch Modeling                & 62.395 & -0.216 \\\hline
Architecture Pathway Ablation               & 50.641 & -11.970 \\\hline
FFT Normalization                     & 62.041 & -0.570 \\\hline
Discrete Wavelet Transform                  & 59.529 & -3.082 \\\hline
Discrete Cosine Transform                   & 58.039 & -4.573 \\
\hline
\end{tabular}
\label{tab:ablationstudy}
\end{table}
% ----------------------------------------------------------------------

%----------------------------------------------------------------------------
%----------------------------------------------------------------------------

\section{Conclusion}
We propose HFEMCNet, a novel compact hybrid frequency-enriched multi-channel network for automatic modulation classification (AMC). By jointly capturing spatial and temporal dependencies and leveraging FFT-magnitude features, HFEMCNet achieves superior classification performance compared to contemporary state-of-the-art deep learning models. Extensive evaluations on RML2016.10a, RML2016.10b, and the over-the-air RML2018.01a datasets demonstrate its robustness and generalization across both simulated and real-world scenarios. Comprehensive computational analysis shows that HFEMCNet is highly resource-efficient, with fewer parameters, reduced memory footprint, and lower tail-latency, making it suitable for deployment on resource-constrained platforms. Ablation studies further reveal that multi-branch architectures, optimal spectral resolution, and careful selection of frequency-domain representations critically impact classification accuracy. In particular, FFT-derived features outperform alternatives such as discrete wavelet transform (DWT) and discrete cosine transform (DCT), highlighting their discriminative effectiveness for AMC. Proposed HFEMCNet combines high classification accuracy with practical deployability, providing a robust and efficient solution for real-world modulation classification tasks.

% %-------------------------------------------------
% \begin{center}
% {\color{red}\rule{0.8\columnwidth}{0.5mm}}
% \end{center}
% %-------------------------------------------------

% or


%\appendices

%\section{}

%\section*{Acknowledgment}


\ifCLASSOPTIONcaptionsoff
  \newpage
\fi



%\bibliographystyle{IEEEtran}
% References
%------------------------------------------------------------------------------

\bibliographystyle{IEEEtran}
\bibliography{references.bib}    % <-- no “.bib” extension

\begin{IEEEbiography}[{\includegraphics[width=1in,height=1.25in,clip,keepaspectratio]{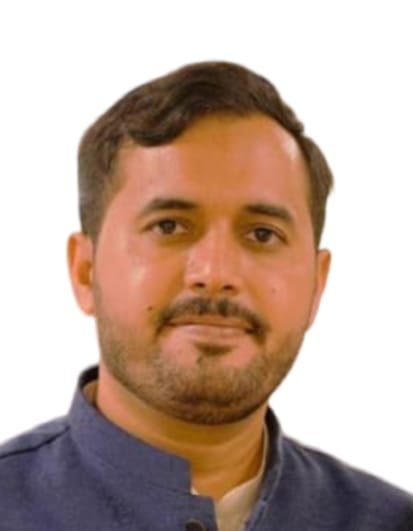}}]{Qamar Ijaz}
received the
bachelor’s degree in avionics engineering from the
National University of Sciences and Technology,
Pakistan, in 2016, where he is currently pursuing
the master’s degree in avionics engineering. His
current research interests include deep learning,
radio signal classification, and artificial intelligence.
\end{IEEEbiography}

\begin{IEEEbiography}[{\includegraphics[width=1in,height=1.25in,clip,keepaspectratio]{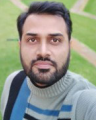}}]{Nayyer Aafaq}
received the B.E. degree
(Hons.) in Avionics Discipline from the College of Aeronautical Engineering (CAE), National University
of Sciences and Technology (NUST), Pakistan,
in 2007; the M.S. degree (Hons.) in systems engineering from the Queensland University of Technology (QUT), Australia, in 2012; and the Ph.D.
degree from the School of Computer Science
and Software Engineering (CSSE), University
of Western Australia (UWA), in 2021. He is currently adjunct faculty as an assistant professor with NUST. His research has won national and international awards. He won the Outstanding Thesis Award in his PhD (0.7\% award rate). He is also a nominee for Australia's prestigious Robert Street Award. He is also the recipient of the SIRF scholarship at UWA, Australia. His research on deep learning-based automated FOD detection won the first consolation prize nationwide out of 148 industry/academia participants. 
His research in computer vision and
pattern recognition has been published in prestigious venues of the field,
including IEEE Computer Vision and Pattern Recognition (IEEE CVPR), IEEE
Transactions on Multimedia (IEEE TMM), IEEE Transactions on  Artificial Intelligence (IEEE TAI),
IEEE Transactions on Information Forensics and Security (IEEE TIFS), and ACM Computing Surveys (ACM CSUR). His current research interests include Deep
Learning, Pattern Recognition, and the intersection of Natural Language Processing (NLP), Computer Vision (CV), and Artificial Intelligence. \end{IEEEbiography}


%\newpage



%\vfill



\end{document}